\documentclass[aps,prd,twocolumn,longbibliography,reprint,superscriptaddress,amsmath,nofootinbib]{revtex4-1}
\usepackage[colorlinks=true, urlcolor=blue,linkcolor=blue, anchorcolor=blue, citecolor=red]{hyperref}
\usepackage[utf8]{inputenc}
\usepackage{natbib}
\usepackage{graphicx}
\usepackage{xcolor}
\usepackage{amsmath}
\usepackage{bm}
\graphicspath{{figs/}}

\usepackage{amsmath,amssymb,amsfonts}
\usepackage{graphicx}
\usepackage{color}
\usepackage{floatflt}
\usepackage{longtable}
\usepackage{dcolumn}
\usepackage[normalem]{ulem}

\newcommand{\dif}{{\rm d}}
\allowdisplaybreaks
\newcommand{\be}{\begin{equation}}
\newcommand{\ee}{\end{equation}}
\newcommand{\bea}{\begin{eqnarray}}
\newcommand{\eea}{\end{eqnarray}}

\begin{document}

\title{Impact of modified gravity theory on neutron star and nuclear matter properties}

\author{Naosad Alam}
\email{naosad.alam@tifr.res.in}
\author{Subrata Pal}
\affiliation{Department of Nuclear and Atomic Physics, Tata Institute of Fundamental Research, Mumbai 400005, India}
\author{A. Rahmansyah}
\author{A. Sulaksono}
\affiliation{Departemen Fisika, FMIPA, Universitas Indonesia, Depok 16424, Indonesia}


\begin{abstract} 
New observational data measured with high degree of accuracy of compact isolated neutron stars and binary stars in gravitational wave remnants have the potential to explore the strong field gravity.
Within the framework of energy-momentum squared gravity (EMSG) theory we study its impact on several properties of neutron stars and plausible modifications from the predictions of general relativity.
Based on a representative set of relativistic nuclear mean field models, non-relativistic Skyrme-Hartree-Fock models and microscopic calculations, we show deviations  of neutron star mass-radius sequence in EMSG theory as compared to general relativity. The variation in the effective
nuclear equation of state in EMSG, results in distinct magnitudes in the reduced pressure, speed of sound, and maximum compactness at the center of neutron stars. We perform extensive correlation analysis of the nuclear model parameters with the neutron star observables in light of the new 
observational bounds. Perceptible modifications in the correlations are found in the 
models of gravity that provide different estimates of the slope and curvature of nuclear matter symmetry energy. The available neutron star data however do not impose stringent enough constraints for a clear evidence on deviations from general relativity.

\end{abstract}


\maketitle
\section{Introduction}
\label{section:intro}

Understanding the stellar structures, such as the compact neutron stars, relies entirely 
on the physics of high-density matter \cite{Lattimer04}. The two major impediments to a precise determination 
of neutron star (NS) properties at supranuclear densities are the lack of detailed knowledge of 
nuclear interaction in particular \cite{Dutra14,Oertel2016,Lattimer14} and gravitational interaction 
\cite{Capozziello2011,NR2018,ABCEK2018,Olmo2019,ABU2020}. The repulsive nuclear equation
of state (EoS: characterizing the dependence of matter pressure $P$ on energy density $\rho$) and the 
balancing attractive strong-field gravitational physics are intertwined via the 
Tolman-Oppenheimer-Volkoff (TOV) equations \cite{Tolman39,Oppenheimer39} for hydrostatic equilibrium of the star configuration, hence their uncertainties could impact the predictions of structure and properties of neutron stars.

While terrestrial experiments and {\it ab inito} calculations provide nuclear matter description 
only about the saturation densities, one relies on several sophisticated nuclear many-body 
interaction theories \cite{Dutra14,Oertel2016} for high density behaviour. These models by 
construction reproduce the ground state nuclear matter properties, 
as a consequence, the higher density predictions of these EOSs are very diverse and remain largely 
unconstrained. Particularly uncertain are the supranuclear density behavior of nuclear symmetry energy $e_{\rm sym}$,
its slope and curvature at saturation density and thus the EoS of neutron-rich matter \cite{Brown00,Li08,Lattimer04}. Considerable attempts have been made to put stringent constraints on the EoS 
by employing the combined measurements of neutron star masses and radii, and the observed tidal
deformability bound from the detected gravitational waves \cite{Yagi13}.

On the other hand, the impact of various theories of gravity in the strong-field regime remains largely 
unexplored \cite{Capozziello2011,Olmo2019}. While Einstein's General theory of Relativity (GR) continues to be a very effective theory 
of gravitational interaction at various scales, especially with the detection of gravitational waves,
sufficient motivation to investigate alternate viable theories of gravity arises from unexplained
dark matter and dark energy at the galactic and cosmological scales, and the presence of 
singularity in the early universe and inside black holes \cite{NR2018}. Being a superdense object with strong gravitational field, neutron stars offer an exciting avenue for investigation of general relativity in the strong field or high curvature domain and open up a direction for the study of new gravitational physics. Hence, it will be appealing to explore and test alternative theories of gravity in the case of superdense stars in addition to the traditional approach based on GR.

In the absence of a fundamental quantum gravity theory for the description of complete gravitational action, the formulation of modified theories mainly focused initially on the simplest 
modification of the gravitational Lagrangian by some analytic function $f({\cal R})$ different from the linear function of the spacetime curvature (the Ricci
scalar) ${\cal R}$.
This approach enforces a modification on the left-hand side of Einstein's field equations 
$\mathcal{R}_{\mu\nu} - g_{\mu\nu}\mathcal{R}/2 = \kappa T_{\mu\nu}$, in the usual notation \cite{Capozziello2011}. Subsequently, a generalized
$f({\cal R, T})$ gravity was developed \cite{Harko11} that also includes the trace ${\cal T}$ of the energy-momentum tensor $T^{\mu\nu}$ of  matter.
More recently, a more generic covariant form was proposed that considers a nonlinear Lorentz scalar involving the entire matter Lagrangian
$f({\cal R}, T^2) \equiv {\cal R} + \alpha (T^2)^n = {\cal R} + \alpha (T^{\mu\nu}T_{\mu\nu})^n$ and dubbed as energy-momentum powered gravity (EMPG), where 
$\alpha$ is a parametric constant
\cite{Katirci14, NR2018,ABCEK2018,ABU2020,Nazari2022,NRM2022,Fazlollahi2023}. This theory provides non-minimal matter and geometry coupling, and introduces a higher-order contribution to the material stresses on the right-hand side of the Einstein's equations.
The case with dimensionless parameter $n=1$ (referred to as energy-momentum squared gravity: EMSG) can be effective at high energy densities relevant to the neutron star interior \cite{ABCEK2018,NR2018}.
The EMSG theory also resolves the ineffectiveness of the $f({\cal R, T})$ gravity for a perfect
fluid with EoS $P = \rho/3$ where the trace ${\cal T} = 0$ and $f({\cal R, T})$ 
reduces to the usual $f({\cal R})$ theory.
The EMSG gives deviations from GR also at low curvature domains \cite{DDAS2023} and compatible to GR at vacuum. 
 Technically, the field equations alter the expressions of the TOV equations, and hence the astrophysical properties of stars.

The EMPG theory suggests a bounce in the early universe due to maximum energy density and correspondingly a minimum length scale factor, thereby addressing the problem of big-bang singularity as well as the current cosmic accelerated expansion. 
This theory also correctly predicts cosmic behaviour and follows the actual progression of cosmological eras \cite{Roshan16}. 
Since EMSG is proposed to resolve the singularities classically, it is expected 
that the deviations from GR appear in the properties of compact stars \cite{NR2018}. 
Further, the  EMSG model passes the weak field tests for the Solar System regime 
where the EMSG light deflection, Shapiro time delay and gravitational microlensing scenarios are found similar to GR \cite{Akarsu22a,Nazari2022}. Moreover, this theory is consistent with the strong-field gravity test as investigated from analysis of the first time derivative of the orbital period of binary pulsars \cite{Nazari2022}. 
Consequently, the EMSG theory is considered to be a viable model that explains the cosmological behavior 
as well as passes the weak and strong gravity-field tests, and thus ideally suited to test GR modifications in neutron stars.

Recently, the parameter $\alpha$ in the EMSG model has been constrained 
\cite{ABCEK2018} by using the $2M_\odot$ maximum mass NS constraint and physicality of certain effective EoSs
in the center of NS, to be in the range $\rm -10^{-38} \lesssim \alpha \lesssim 10^{-37} ~ cm^3/erg$. 
Whereas, binary pulsar observation \cite{ABU2020} yields a compatible value of $\rm -6 x 10^{-38} \lesssim \alpha \lesssim 10^{-36} cm^3/erg$, but that reported from solar system test are relative lower 
$ \rm -4 x 10^{-27}  < \alpha < 10^{-26} ~ cm^3/erg$. 

In the present study, we shall compare the predictions of EMSG theory relative to general relativity to explore the strong field effects in neutron stars.
The field equations in EMSG with the standard physical energy-momentum 
tensor can be mapped into the GR Einstein's field equations, but with an effective or modified energy-momentum tensor. This enables a rather straightforward calculation of the moment of inertia, tidal deformation, and other properties of the stars \cite{RPKS2022}.

For the nuclear EoS, we have employed a comprehensive set of relativistic mean field (RMF) theory 
for nuclear interaction that 
provides Lorentz covariant extrapolation from sub- to supra-saturation densities.
The model has been extensively applied in the description of several finite nuclei properties and
studies in NS structure. Further, we have employed a representative set of non-relativistic 
Skyrme-Hartree-Fock (SHF) model, and two microscopic 
theories based on Brueckner-Hartree-Fock (BHF) and variational approaches. 
Within these model EoSs, that have diverse high-density behavior, we shall explore the EMSG 
and the GR effects on the neutron star properties. Furthermore, we will examine the correlations 
between the NS properties, namely mass, radius, tidal deformability with the key nuclear EoS 
parameters (as well as between the thermodynamic variables, namely the pressure, speed of sound). 
Only if tight correlations between NS observables and EoS parameters in various models of gravity
can be established, one can then provide suitable (model independent) bounds on 
these nuclear matter quantities by employing the precisely measured NS observables. 
Alternatively, these relations can be used to constrain an astrophysical observable 
from the knowledge of the correlated nuclear matter observables.
In fact, by using a larger number of unified EoSs, extensive correlation analysis in general relativity
have been conducted between neutron star mass $M$, radius $R$, etc with the parameters 
of the nuclear EoS, such as the nuclear matter incompressibility $K(\rho_0)$, 
its slope $M(\rho_0)$, the nuclear symmetry energy slope $L(\rho_0)$ and 
curvature $K_{\rm sym}(\rho_0)$,
at the saturation density $\rho_0 \approx 0.16$ fm$^{-3}$ \cite{Fortin16}, and their 
linear combinations \cite{Alam16,Malik18,Carson19,Burgio21,Yang23},
as well as with the tidal deformability 
$\Lambda$ \cite{Malik18} and corrections to mass-weighted tidal deformability \cite{Carson19} of the 
detected gravitational waves GW170817
\cite{LIGO2017vwq}. While the individual EoS parameters were found to be weakly
correlated, their specific linear combinations showed a rather strong correlation \cite{Yagi13,Alam15,Alam16,Malik18,Carson19,Yang23}. It will be
instructive to investigate and understand 
how these correlations between the astrophysical observables and nuclear EoS behave in the 
alternative energy-momentum squared gravity (EMSG) model as compared to the predictions of 
General Relativity, and whether an approximate universal constraint can be imposed on the EoS parameters 
that is independent of nuclear and gravitational interactions.

The outline of the paper is as follows. In Sec.~\ref{sec:EMSG}, we briefly describe the modified field equations in EMSG. The modified structure in TOV equations for neutron star in EMSG are discussed in Sec.~\ref{sec:hydrostatics}. We then discuss the methodology to compute the moment of inertia in slow rotation approximation in Sec.~\ref{sec:MI} and the tidal deformability parameter in Sec.~\ref{sec:TD}. Next, we provide a brief review of the key EoS parameters and the EoSs used the analysis in Sec.~\ref{sec:eos}. Our results on the calculations of 
neutron star configurations within EMSG and GR are presented in Sec.~\ref{sec:results}. Within various diverse EoSs, correlations 
between the parameters of the EoSs and the NS properties in the EMSG 
modified theory of gravity will be also discussed.
Finally, the conclusions are drawn in Sec.~\ref{sec:conclusion}. 
We will adopt the system of units $\hbar=c=G=1$ throughout the manuscript.

\section{Energy-Momentum Squared Gravity}
\label{sec:EMSG}

In the energy-momentum squared gravity theory, the Einstein-Hilbert action is modified by the addition of a scalar term $f(T_{\mu\nu}T^{\mu\nu}) = \alpha T_{\mu\nu}T^{\mu\nu}$ leading to \cite{ABCEK2018,ABU2020,Olmo2019}:
\begin{align}
S=\int \left[\frac{1}{2\kappa}\left(\mathcal{R}-2\Lambda\right)+\alpha T_{\mu\nu}T^{\mu\nu}+ \mathcal{L}_{\mathrm{m}}\right]\sqrt{-g}\,\mathrm{d}^4x,
\label{action}
\end{align}
where $\kappa =8\pi G$ is Newton's constant, $\mathcal{R}$ denotes the Ricci scalar, 
$g$ is the determinant of the metric, $\Lambda$ the cosmological constant, and $\alpha$ the coupling parameter. 
The Lagrangian density $\mathcal{L}_{\mathrm{m}}$ represents source of the matter
described by the energy-momentum tensor which can be defined as usual:
\begin{align}  \label{tmunudef}
T_{\mu\nu}=-\frac{2}{\sqrt{-g}}\frac{\delta(\sqrt{-g}\mathcal{L}_{\mathrm{m}})}{\delta g^{\mu\nu}}=g_{\mu\nu}\mathcal{L}_{\mathrm{m}}-2\frac{\partial \mathcal{L}_{\mathrm{m}}}{\partial g^{\mu\nu}}.
\end{align}
Consequently, the Einstein's field equation for the modified action becomes
\begin{align}
G_{\mu\nu}+\Lambda g_{\mu\nu}=\kappa T_{\mu\nu}+\kappa \alpha \left(g_{\mu\nu}T_{\sigma\epsilon}T^{\sigma\epsilon}-2\theta_{\mu\nu}\right),
\label{fieldeq}
\end{align}
where $G_{\mu \nu }=\mathcal{R}_{\mu \nu }-\frac{1}{2}g_{\mu\nu}\mathcal{R}$ is the Einstein tensor and the new tensor $\theta _{\mu\nu}$ is defined as
\begin{align} 
\theta_{\mu\nu}=& T^{\sigma\epsilon}\frac{\delta
T_{\sigma\epsilon}}{\delta g^{\mu\nu}}+T_{\sigma\epsilon}\frac{\delta
T^{\sigma\epsilon}}{\delta g^{\mu\nu}} \nonumber \\ 
=&-2\mathcal{L}_{\mathrm{m}}\left(T_{\mu\nu}-\frac{1}{2}g_{\mu\nu}{\cal T}\right)- 
{\cal T} T_{\mu\nu}  \nonumber \\
&+2T_{\mu}^{\gamma}T_{\nu\gamma}-4T^{\sigma\epsilon}
\frac{\partial^2\mathcal{L}_{\mathrm{m}}}{\partial g^{\mu\nu} \partial g^{\sigma\epsilon}}.
\label{theta} 
\end{align}
Here ${\cal T}=g^{\mu \nu}T_{\mu \nu}$ is the trace of the energy-momentum tensor.
We consider the star to be a perfect fluid (i.e. non-viscous and stress-free), 
with energy-momentum tensor $T_{\mu\nu}=(\rho+P)u_{\mu}u_{\nu}+P g_{\mu\nu}$,
where $\rho$ is the energy density, $P$ is the isotropic pressure, and $u_{\mu}$ is the four-velocity.
Since the definition of matter Lagrangian for the perfect fluid described via the energy-momentum tensor is not unique, one can consider $\mathcal{L}_{\rm m}=P$ or $\mathcal{L}_{\rm m}=-\rho$; both these choices lead to the same $T^{\mu\nu}$ in the case of GR. In contrast, for non-minimal coupling of matter with gravity as in EMSG, it gives rise to distinct theories with different predictions \cite{Haghani23,Akarsu23}.
In this work we consider the former choice of $\mathcal{L}_{\rm m}=P$ that has been commonly employed to construct a viable astrophysical/cosmological model ~\cite{Haghani23,Akarsu23,Faraoni09}. The covariant divergence of Eq. \eqref{fieldeq} then becomes 
\begin{equation}
\nabla^{\mu}T_{\mu\nu}=-\alpha g_{\mu\nu}\nabla^{\mu}
(T_{\sigma\epsilon}T^{\sigma\epsilon})+2\alpha\nabla^{\mu}\theta_{\mu\nu},
\label{nonconservedenergy}
\end{equation}
Note that the local/covariant energy-momentum conservation $\nabla^{\mu}T_{\mu\nu}$ is not identically zero for $\alpha\neq 0$.
Using Eqs. \eqref{fieldeq}, \eqref{theta} and the above definition of $T_{\mu\nu}$, 
one finally obtains \cite{ABCEK2018}
\begin{align}
&G_{\mu\nu}+\Lambda g_{\mu\nu}=\kappa \rho \left[\left(1+\frac{P}{\rho}\right)u_{\mu}u_{\nu}+\frac{P}{\rho}g_{\mu\nu}\right]\nonumber \\ 
&+\alpha\kappa\rho^2\left[2\left(1+\frac{4P}{\rho}+\frac{3P^2}{\rho^2}\right)u_{\mu}u_{\nu}+\left(1+\frac{3P^2}{\rho^2}\right)g_{\mu\nu}\right]. 
\label{fieldeq2}
\end{align}
Equation \eqref{fieldeq2} can be recast into GR Einstein's field equation
\begin{align}
G^{\mu\nu}+\Lambda g^{\mu\nu}=\kappa T^{\mu\nu}_{\rm eff},
\label{fieldeq3}
\end{align}
with an effective energy momentum tensor $T^{\mu\nu}_{\rm eff} 
=(\rho_{\rm eff}+P_{\rm eff})u^{\mu}u^{\nu}+P_{\rm eff} g^{\mu\nu}$ for an ideal fluid,
where the effective energy density and pressure are given by
\begin{align} \label{eqn:f1}
\rho_{\rm eff} & = \rho+\alpha\rho^2\left(1+\frac{8P}{\rho}+\frac{3P^2}{\rho^2}\right), \\
P_{\rm eff} & = P+\alpha\rho^2\left(1+\frac{3P^2}{\rho^2}\right).
\label{eqn:f2}
\end{align}
It is important to note that the EMSG model with an isotropic energy-momentum tensor
(ideal fluid) can be mapped exactly into GR with an {\rm isotropic} effective
$T^{\mu\nu}_{\rm eff}$ appearing only in the material side of the field equations \cite{Akarsu23}. 
While many modified gravity theories (such as $f({\cal R})$) do not have this feature, the Eddington-inspired-Born-Infeld (EiBI) theory although can be mapped into 
GR but has an {\rm anisotropic} effective energy-momentum tensor \cite{DDAS2023,RPKS2022}. 
Hence, the mapped expressions for the field equations in case of EMSG theory allows a straightforward calculation of the NS properties.
One can easily see from Eq. \eqref{fieldeq3} that twice contracted Bianchi identity yields vanishing of the covariant divergence
$\nabla_\mu T^{\mu\nu}_{\rm eff} = 0$.
Thus, the (isotropic) effective energy-momentum tensor is conserved in EMSG gravity. This generalized conservation is local (stemming from equivalence principle), which is the actual criterion that should be considered for modified gravity and not the conservation
$\nabla_\mu T^{\mu\nu} = 0$ of conventional matter fluids. These equations in curved spacetime have essentially distinct description from the usual conservation $\partial_\mu T^{\mu\nu} = 0$ 
in absence of gravity; detail discussion can be found in Ref. \cite{Khodadi:2022zyz}.
Consequently, due to isotropy of the effective energy-momentum tensor and its conservation in EMSG, one does not require any additional equations for thermodynamical consistency via the general maximum entropy principle \cite{Gao11,*Gao12,Wojnar23}.

 \section{ToV equations in EMSG}
\label{sec:hydrostatics}

To obtain the Tolman-Oppenheimer-Volkoff equations \cite{Tolman39,Oppenheimer39} for a non-rotating star 
in the EMSG description, we adopt the general spherically symmetric metric as
\begin{equation}  \label{eqn:metric}
\mathrm{d} s^2 = -e^{2\nu\left(r\right)}\mathrm{d} t^2 +e^{2\lambda\left(r\right)}\mathrm{d} r^2+r^2\mathrm{d}\theta^2+r^2\sin^2\theta \, \mathrm{d}\phi^2 ,
\end{equation}
where metric functions $\nu(r)$ and $\lambda(r)$ depend only the radial coordinate $r$.
Using Eqs. \eqref{fieldeq} and \eqref{eqn:metric}, one obtains the $(tt)$ and $(rr)$ components of EMSG field equation as
\begin{align} \label{eqn:mf1}
\frac{1}{r^2}-\frac{e^{-2\lambda}}{r^2}\left(1-2r\frac{\dif \lambda}{\dif r}\right) & = \kappa\rho_{eff},
\\
-\frac{1}{r^2}+\frac{e^{-2\lambda}}{r^2}\left(1+2r\frac{\dif\nu}{\dif r}\right) & =\kappa P_{eff},  
\label{eqn:mf2}
\end{align}
where $\rho_{\rm eff}$ and $P_{\rm eff}$ are the effective values of mass density and pressure 
at a distance $r$ from the center of NS. 
By defining the metric function $\lambda\left(r\right)$ in terms of the mass function $m(r)$ as
\begin{equation} \label{eqn:pt1} 
e^{-2\lambda\left(r\right)}=1-\frac{2m\left(r\right)}{r}, 
\end{equation} 
and the metric function $\nu (r)$ via the pressure as \cite{ABCEK2018}
\begin{align}  \label{eqn:pt2}
\frac{\mathrm{d} \nu}{\mathrm{d} r} =&- \left[\rho \left( 1+\frac{P}{\rho}\right) \left\{ 1+2\alpha\rho\left( 1+\frac{3P}{\rho} \right)\right\} \right]^{-1}  \notag \\
&\times \left[ \left(1+6\alpha P\right) \frac{\mathrm{d} P}{\mathrm{d} r}+2\alpha\rho\frac{\mathrm{d}\rho}{\mathrm{d} r} \right],
\end{align} 
one obtains the modified TOV equations in EMSG:
\begin{align} \label{TOV1}
\frac{\mathrm{d} m}{\mathrm{d} r} = & 4\pi r^2 \rho \left[1+\alpha\rho \left(1+\frac{8P}{\rho}+\frac{3P^2}{\rho^2 }\right)\right],   \\
\frac{\mathrm{d} P}{\mathrm{d} r}  = & -\frac{m\rho }{r^2}\left(1+\frac{P}{\rho}\right) \left( 1-\frac{2m}{r}\right)^{-1}  \nonumber \\
&\times \left[ 1+\frac{4\pi r^3 P}{m }+\alpha \frac{4\pi r^3\rho^2}{m}\left(1+\frac{3P^2}{\rho^2}\right)\right] \notag \\
&\times \left[1+2\alpha\rho\left(1+\frac{3P}{\rho} \right)\right] \left[1 + 2\alpha\rho \left(\frac{\dif \rho}{\dif P}+\frac{3P}{\rho}\right)\right]^{-1}, 
\label{TOV2}
\end{align}
The structure of the relativistic stars i.e. the mass and radius can be obtained by solving Eqs. \eqref{TOV1}-\eqref{TOV2} simultaneously with an input EoS $P \equiv P(\rho )$, which describes the 
relation between the pressure $P(r)$ and the density $\rho(r)$ of the matter.
It is evident from Eqs. \eqref{TOV1}-\eqref{TOV2} or equivalently from 
 Eqs. \eqref{eqn:f1}-\eqref{eqn:f2}, that EMSG modifications
to GR for the NS configurations stem from
the additional terms contributing to the energy density 
$\rho_{\rm EMSG} = \alpha(\rho^2 + 8\rho P + 3P^2)$
and pressure $P_{\rm EMSG} = \alpha(\rho^2  + 3P^2)$.
Due to isotropic and perfect fluid nature of the effective energy-momentum tensor, one can also obtain the above generalized TOV equations by using the effective thermodynamic variables and maximizing the effective
entropy \cite{Gao11,*Gao12,Wojnar23}.

\section{Moment of Inertia}
\label{sec:MI}

In this section we briefly present the calculation of the moment of inertia of a rotating neutron star in the energy-momentum squared gravity. We consider that the star rotates uniformly with a stellar frequency $\Omega$ which is much lower in comparison with the Kepler frequency at the equator, i.e.
$\Omega \ll \Omega_{\rm max} \approx \sqrt{M/R^3}$.
The moment of inertia of such an axially symmetric and uniformly rotating neutron star \cite{Hartle67,Morrison04} in EMSG can be written as
\begin{equation}
  I \equiv \frac{J}{\Omega} = \frac{8\pi}{3}
 \int_{0}^{R} r^{4} e^{-\nu(r)}\frac{\bar{\omega}(r)}{\Omega}
 \frac{\left[\rho_{\rm eff}(r)+P_{\rm eff}(r)\right]}{\sqrt{1-2m(r)/r}} dr .
 \label{MomInertia}
\end{equation}
Note that the effective energy density $\rho_{\rm eff}$ and pressure $P_{\rm eff}$ of Eqs. \eqref{eqn:f1} and \eqref{eqn:f2} enter the expression. $J$ is the angular momentum, $\nu(r)$ and $\bar{\omega}(r)$ are the metric functions.
In the slowly rotating approximation, the line element for the background metric of a stationary and axially symmetric star can be taken as
\begin{align}
   ds^2_r = & -e^{2\nu(r)}dt^2 + e^{2\lambda(r)} dr^2 +
  r^2 d\theta^2\nonumber \\
  &+ r^2 \sin^2\theta d\phi^2 - 
 2\omega(r)r^2\sin^2\theta dt d\phi .
 \label{metric}  
\end{align}
Here the metric functions $\nu(r)$ and $\lambda(r)$ will be identical to the case of a static and spherically symmetric neutron star, and simply follow  Eqs. \eqref{eqn:pt1} and \eqref{eqn:pt2}.

To calculate the moment of inertia, we further require the form of metric function $\omega(r)$ which appears due to the slow rotation of the star. The dimensionless relative frequency, defined as
 \begin{equation}
 \bar{\omega}(r)\!\equiv\!\frac{\Omega-\omega(r)}{\Omega} ,   
 \end{equation}
 obeys the differential equation
\begin{equation}
 \frac{d}{dr}\left[r^{4}j(r)\frac{d\bar{\omega}(r)}{dr}\right]
 +4r^{3}\frac{dj(r)}{dr}\bar{\omega}(r) = 0 ,
 \label{OmegaBar}
\end{equation}
where 
\begin{equation}
  j(r)=e^{-\nu(r)-\lambda(r)} = 
  \begin{cases} 
    e^{-\nu(r)}\sqrt{1-2m(r)/r}  & \text{if $r \le R ,$}\\
    1 &\text{if $r > R.$}
  \end{cases}
\end{equation}
The solution to the above equation can be obtained by using the following two boundary conditions:
\begin{subequations}
 \begin{align}
  & \bar{\omega}'(0)=0 , 
  \label{BC1}\\
  & \bar{\omega}(R)+\frac{R}{3}\,\bar{\omega}'(R)=1.
  \label{BC2}
 \end{align}
\end{subequations}
To solve the differential equation \eqref{OmegaBar}, one can start with a guess value 
of the central frequency $\bar{\omega}_{c}\!=\!\bar{\omega}(0)$ and numerically
integrate the equation up to the surface of the star. Since we start with an arbitrary
value of $\bar{\omega}_{c}$, usually, the boundary condition at $R$ will not be 
satisfied. However, this can be achieved by simply rescaling $\bar{\omega}_{c}$ by an 
appropriate constant. Once we have the solution of $\bar{\omega}(r)$, the moment of 
inertia can be calculated from Eq. \eqref{MomInertia}. After obtaining the solutions 
of $\bar{\omega}(r)$ and $I$, the consistency of the formalism may be 
verified from
the condition $\bar{\omega}'(R) = 6GI/R^{4}$ \cite{Hartle67,Morrison04}.

\section{Tidal Deformability}
\label{sec:TD}

The phase of the gravitational wave signal resulting from the merger of two neutron stars carries valuable information about the tidal deformability parameter that is directly related to the internal structure and composition of the star, particularly the equation of state of nuclear matter. It quantifies the deformations induced in the star due to an external tidal field of the companion star. The tidal deformability parameter $\lambda$ can be expressed as \cite{Flanagan08,Hinderer08,Hinderer10,Damour12,Alam15,Malik18},
\begin{equation}
\lambda = - \frac{Q_{ij}}{ {\cal E}_{ij}},
\end{equation}
where $Q_{ij}$ represents the components of the induced quadrupole moment tensor and ${\cal E}_{ij}$ denotes the components of the tidal field tensor. In terms of the Love  number $k_2$, the  mass normalized dimensionless tidal deformability parameter is given by
\begin{equation}
\Lambda \equiv \frac{\lambda}{M^{5}} = \frac{2}{3}k_2\left(\frac{R}{M}\right)^5
\equiv \frac{2}{3}k_2 C^5 ,
\label{eq:Lambda}
\end{equation}
where $R$ and $M$ are the radius and mass of the star, and $C \equiv M/R$ is its compactness.

The tidal Love number $k_2$ depends on the underlying EoS of the star and it can be expressed in terms of the dimensionless compactness parameter $C$ as~\cite{Flanagan08,Hinderer08,Hinderer10,Damour12},
\begin{align}
k_2 &= \frac{8C^5}{5}\left(1-2C\right)^2
\left[2+2C\left(y_R-1\right)-y_R\right]  \nonumber\\
& \times \big\{ 2C\left[6-3 y_R+3 C(5y_R-8)\right] \nonumber\\
& +4C^3\left[13-11y_R+C(3 y_R-2) + 2 C^2(1+y_R)\right] \nonumber\\
& +3(1-2C)^2\left[2-y_R+2C(y_R-1)\right]\ln\left(1-2C\right) \big\}^{-1}.
\label{eq:k2}
\end{align}
 The function $y_R \equiv y(r)|_{r=R}$ is related to the metric perturbation and satisfies the following differential equation:
\begin{align}
r \frac{d y(r)}{dr} + {y(r)}^2 + y(r) F(r) + r^2 Q(r) = 0
\label{TidalLove2} ,
\end{align}
where the functions $F(r)$ and $Q(r)$ are given
\begin{align}
F(r) &= \frac{r-4 \pi r^3 \left[ \rho_{\rm eff}(r) - P_{\rm eff}(r)\right] }{e^{-2\lambda\left(r\right)}} ,
\nonumber \\
Q(r) & = \frac{4\pi}{e^{-2\lambda\left(r\right)}} \Big[ 5\rho_{\rm eff}(r) +9 P_{\rm eff}(r) +
\frac{\rho_{\rm eff}(r) + P_{\rm eff}(r)}{\partial P_{\rm eff}(r)/\partial\rho_{\rm eff}(r)}
\nonumber \\
&- \frac{6}{4 \pi r^2}\Big] 
-4\left[\frac{m(r) + 4 \pi r^3 P_{\rm eff}(r)}
{r^2 e^{-2\lambda\left(r\right)}}\right]^2 .
\label{Qr}
\end{align}
 For a spherically symmetric star, the Love number and tidal deformability parameter $\Lambda$ can be determined by simultaneously solving Eq. (\ref{TidalLove2}) and the TOV equations (Eqs. \eqref{TOV1}-\eqref{TOV2}), with the boundary conditions $P(0)\!=\!P_{c}$ and $m(0)\!=\!0$ in addition to $y(0) = 2$ that arises from perturbative expansion of the deformed metric up to the second order.

\section{Nuclear matter equations of state}
\label{sec:eos}

In the parabolic approximation, the equation of state of isospin asymmetric nuclear matter at a given density $\rho$ and asymmetry $\delta$ can be written as \cite{Li08,Tsang12}
\begin{equation}
 e(\rho,\delta) = e_0(\rho) + e_{\rm sym}(\rho)\delta^2 + \mathcal{O}(\delta^4) ,\\
 \label{eq:eden}
\end{equation}
where $e(\rho,\delta)$ is the total energy per nucleon at a nucleon density $\rho=\rho_n+\rho_p$, and  $\delta=(\rho_n-\rho_p)/\rho$ is neutron-proton asymmetry parameter, with $\rho_n$ and $\rho_p$ the neutron and proton densities, respectively. The first term on the right-hand side 
$e_0(\rho) \equiv e(\rho,\delta=0)$ represents the EoS for symmetric nuclear matter, and the second term $e_{\rm sym}(\rho) \equiv \frac{1}{2}{\frac{\partial^2e(\rho,\delta)}{\partial\delta^2}|_{\delta=0}}$ is the nuclear symmetry energy.
The isoscalar part $e_0(\rho)$ and the isovector part $e_{\rm sym}(\rho)$ can be further Taylor series expanded around the saturation density $\rho_0$ as
\begin{align} \label{eq:E}
e_0(\rho) & = e_0(\rho_0)+\frac{K_0}{2} \chi^2 + \frac{Q_0}{6} \chi^3 +\mathcal{O}(\chi^4) , \\
e_{\rm sym}(\rho) & = e_{\rm sym}(\rho_0) + L\chi + \frac{K_{\rm sym}}{2} \chi^2 + \mathcal{O}(\chi^3),
\label{eq:S}
\end{align}
where the dimensionless variable $\chi = (\rho-\rho_0)/3\rho_0$ gives the deviation of density
from the saturation value $\rho_0$. The saturation parameters for the symmetric nuclear matter are the binding energy per nucleon $e_0 \equiv e_0(\rho_0)$,
incompressibility $K_0=9\rho_0^2{\frac{\partial^2e_0(\rho)}{\partial \rho^2}|_{\rho_0}}$,
skewness coefficient $Q_0=27\rho_0^3{\frac{\partial^3 e_0(\rho)}{\partial \rho^3}|_{\rho_0}}$. Similarly, the parameters for the symmetry energy expansion are the symmetry energy coefficient 
$J \equiv e_{\rm sym}(\rho_0)$, and the slope and curvature of symmetry energy i.e.
$L=3\rho_0 {\frac{\partial e_{\rm sym}(\rho)}{\partial \rho}|_{\rho_0}}$ and
$K_{\rm sym}=9\rho_0^2 {\frac{\partial^2 e_{\rm sym}(\rho)}{\partial \rho^2}|_{\rho_0}}$, respectively.

The slope of the incompressibility, 
$M_0=M(\rho_0)=3\rho_0{\frac{\partial K_0(\rho)}{\partial\rho}|_{\rho_0}}$, at the saturation density can be expressed in terms of $Q_0$ and $K_0$ as \cite{Alam15}
\begin{eqnarray}
M_0 = Q_0 + 12K_0 ,     
\end{eqnarray}
and the symmetry energy incompressibility is defined as
$K_{\tau} = 9\rho^2_{\delta} \frac{\partial^2e_{sym}(\rho)}
{\partial \rho^2}|_{\rho_{\delta}}$, where
$\rho_{\delta}$ is the saturation density of asymmetric nuclear matter
corresponding to the asymmetry $\delta$. The symmetry energy 
parameters
$K_{sym}$ and $K_{\tau}$ are related by the following expression \cite{Alam15},
\begin{eqnarray}
K_{\tau}=  K_{sym} - 6L - \frac{Q_0}{K_0}L.
\end{eqnarray}

For analysis of neutron star properties, we employ a representative set of 18 relativistic mean field (RMF) models
\cite{Oertel2016,Dutra14}, 24 non-relativistic Skyrme-Hartree-Fock (SHF)-type models, and 2 microscopic calculations, one of these use the
Brueckner-Hartree-Fock (BHF) approach with Argonne $V_{\rm 18}$ plus 3-body 
Urbana-type nuclear potentials \cite{Taranto13,Davesne16}, and the other
a variational approach namely the Akmal-Pandharipande-Ravenhall (APR) EoS \cite{Akmal98,Ducoin11}.

In the RMF model, the nucleon-nucleon interactions
are described by the exchange of
scalar-isoscalar $\sigma$ meson, vector-isoscalar $\omega$ meson, and vector-isovector $\rho$ mesons. Over the years, the model 
has been improved by the inclusion of non-linear self- and cross-couplings between the mesons.
Based on the form of the interactions in the Lagrangian
density, the RMF models that we have employed in this study, can be broadly classified as: NL-type with nonlinear $\sigma$ term \cite{Centelles90,Liu10}; 
NL3-type with additional $\sigma$-$\rho$ and $\omega$-$\rho$ term, \cite{Lalazissis97},
NL$3{\sigma\rho}4$, NL$3{\sigma\rho}6$ \cite{Pais16}, NL$3{\omega \rho}02$ \cite{Horowitz01}, NL$3{\omega \rho}03$ \cite{Carriere03}; 
TM-type with nonlinear $\omega$ term, TM1 \cite{Sugahara94}, TM1-2 \cite{Providencia13}; FSU-type with an additional form of nonlinear $\omega$ coupling FSU2 \cite{Chen14}; 
BSR-families with more nonlinear couplings \cite{Dhiman07,Agrawal10};
and DD-type with density-dependent couplings,
DD2 \cite{Typel10}, DDH$\delta$ \cite{Gaitanos04}, DDH$\delta$Mod \cite{Ducoin11}, DDME1 \cite{Niksic02}, 
DDME2 \cite{Lalazissis05}, TW \cite{Typel99}, and the GM1
 \cite{Glendenning91}.
 
The SHF models we have taken in the present calculation are SKa, SKb \cite{Kohler76}, SkI2, SkI3, SkI4, SkI5 \cite{Reinhard95}, SkI6 \cite{Nazarewicz96}, Sly2, Sly9 \cite{ChabanatPhd},
Sly230a \cite{Chabanat97}, Sly4 \cite{Chabanat98}, SkMP \cite{Bennour89}, SKOp \cite{Reinhard99},
KDE0V1 \cite{Agrawal05}, SK255, SK272 \cite{Agrawal03}, Rs \cite{Friedrich86}, BSk20, BSk21
\cite{Goriely10}, BSk22, BSk23, BSk24, BSk25, and BSk26
\cite{Goriely13}. 
The coupling constants are obtained
by sophisticated fitting procedures to the finite nuclei
such as the binding energies and charge radii, and the infinite nuclear matter properties at the saturation density $\rho_0$.

All the models considered here have been successful in reproducing various experimental data for finite nuclei. These models are also consistent with $2M_{\odot}$ constraint for the measured maximum mass  of neutron star in general relativity. The SHF-type models can often exhibit a causality problem at very high densities. The SHF models that we have selected in this study, do not become causal up to the central density of neutron star with mass $\sim 2M_{\odot}$. To obtain the EoS for neutron star matter, we have employed a unified inner-crust-core EoS, i.e. the inner crust EoS and the core EoS have been calculated using the same nuclear model, and the outer crust EoS is taken from the work of Baym-Pethick-Sutherland \cite{Baym71}.

The values of the EOS parameters at $\rho_0$ and the corresponding properties of neutron star obtained 
in these models show a significant variation. In this regard, we note that large-scale analysis 
\cite{Oertel2016} of experimental data from finite nuclei and 
heavy-ion collisions with various model calculations
have provided reliable bounds on incompressibility of symmetric nuclear matter 
$210 \leq K_0 \leq 260$ MeV \cite{Colo2004,Colo2013},
the symmetry energy $28\leq e_{\rm sym}(\rho_0) \leq 34$ MeV \cite{Li2019xxz}
from combined analysis of observational data, and a reasonable constraint on the slope of
symmetry energy $46 \leq L\leq 106$ MeV \cite{Fan14,Reed2021,Reinhard2021} at the
saturation density $\rho_0$. However, other nuclear matter
parameters are not constrained and exhibit wide variations even at the saturation
density. The large set of models of different classes employed in the present study 
will predict different NS configurations and thus will allow us to perform the correlation
analysis between the nuclear matter parameters and NS observables with more accuracy.

It may be mentioned that alternative approaches have been developed to construct the high-density  EoS by incorporating state-of-the-art chiral effective field theories
at the low nuclear matter density $\rho \lesssim 1.5\rho_0$ and the quark EoS based on perturbative Quantum Chromodynamics for $\rho \gtrsim 40 \rho_0$. 
The EoS at the intermediate density region is constructed by interpolation methods \cite{Komoltsev22,Lope-Oter22,Lope-Oter23} or piecewise polytropes \cite{Annala20} those satisfy the causality condition of sound speed squared $c_s^2 \leq 1$ and follow thermodynamic consistency. This approach is independent of models of gravity but ignores the crucial nuclear interactions at intermediate densities. On the other hand, we note that the microphysics used to calculate the dense matter EoS
may depend on the inherent gravity theory. Nontrivial effects of curved spacetime have been indicated 
in calculations performed in the Fermi gas EoS by maximizing the entropy in Palatini $f({\cal R})$ gravity 
\cite{Wojnar23} and in the simple $\sigma-\omega$ model by including the effect of gravitational time dilation
\cite{Hossain21}. In contrast, it has been shown
that for hydrostatic equilibrium of dense matter within GR and relativistic fluid dynamics, the EoS should be calculated in flat spacetime, so as to be consistent
with local thermodynamic relations and energy-momentum conservation of the fluid \cite{Li:2022url}.
In what follows, as our study involves diverse sets of realistic EoSs derived from different underlying
microscopic theories with distinct nuclear interactions, such involved calculation incorporating curved spacetime
for various EoSs is beyond the scope of the present paper. Thus, the study mainly focuses on the effects of modified
gravity from matter Lagrangian on the
modification of hydrostatic equilibrium
equations only and thereby the
properties of neutron stars and their connections to the nuclear matter parameters.

\section{Results and Discussions}
\label{sec:results}

In this section, we first discuss with a few selected nuclear EoS how the results in the
EMSG model for gravity differ from GR for the NS configurations due to modifications of the 
hydrostatic equilibrium. Thereafter, we will focus on correlation analysis between the nuclear matter
parameters and properties of neutron stars composed of neutrons, protons, electrons and muons in
$\beta$-equilibrium.

\subsection{Neutron star properties in EMSG theory}

\begin{figure}[t]
 \begin{center}
\includegraphics[width=0.45\textwidth,angle=0]{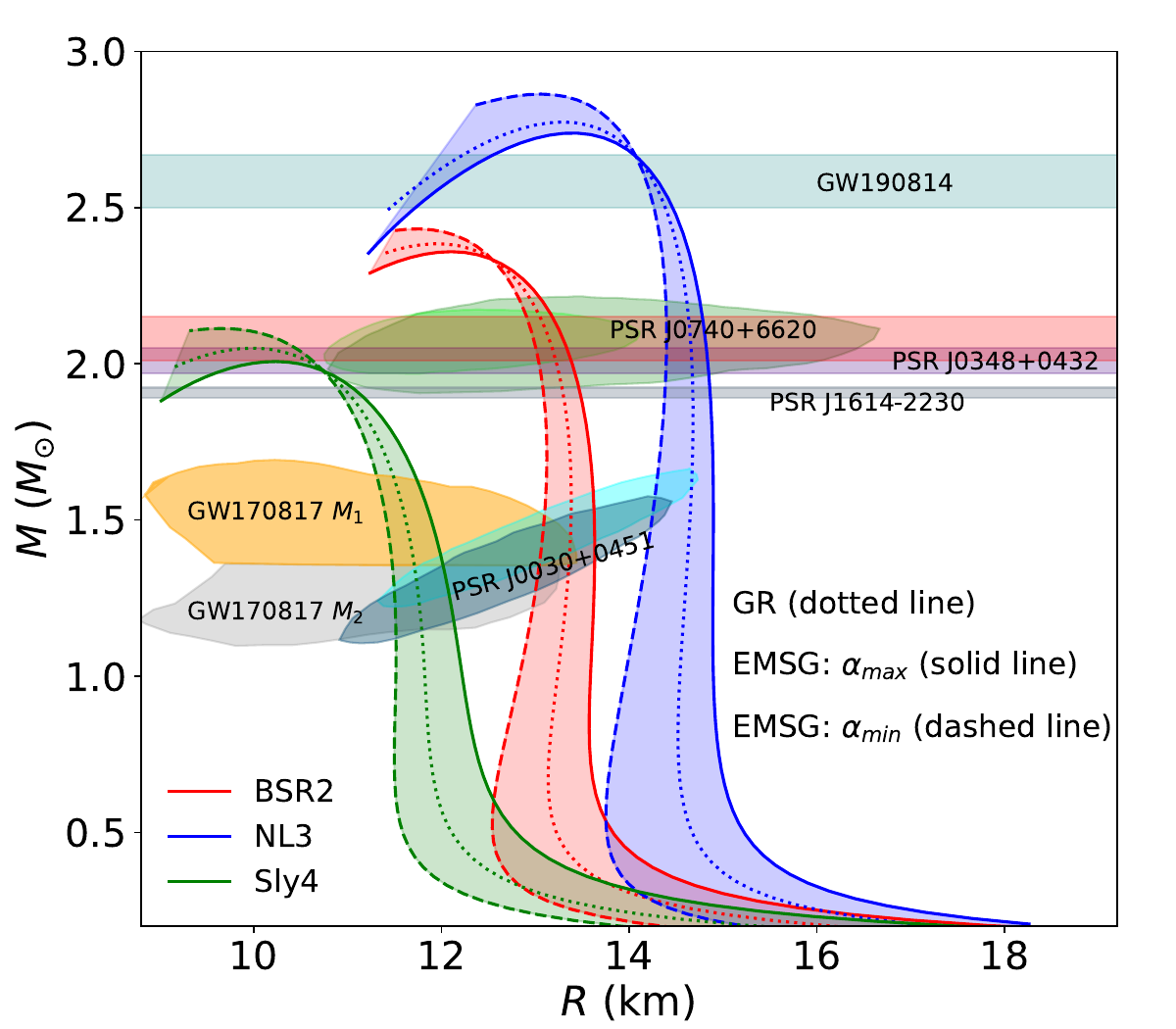}
\end{center}
\vspace{-0.7cm}
  \caption{Neutron star mass-radius curves in the BSR2, NL3, and Sly4 nuclear EoSs. 
  The results are in GR (dotted lines) and in EMSG with a maximum value for $\alpha$ parameter of 
  $\alpha_{\rm max} = 10^{-37}$ cm$^3$/erg (solid lines) and minimum values of
  $\alpha_{\rm min} = -(1.9, 1.2, 0.8)\times 10^{-37}$ cm$^3$/erg (dashed lines) that correspond to stable configuration stars in NL3, BSR2, Sly4 EoSs, respectively (see text for details). The contours and bands refer to $M$-$R$ constraints from NICER measurements of PSR J0030+0451 \cite{Riley2019} and PSR J0740+6620 \cite{Riley2021}, 
the pulsars PSR J0348+0432 \cite{Antoniadis13} and PSR J1614-2230 \cite{Demorest10, Arzoumanian18}, 
the secondary component of the gravitational waves GW190814 
with mass $2.59^{+0.08}_{-0.09} M_\odot$ \cite{LIGOS2020zkf} (horizontal bands), and 
from the GW170817 event \cite{LIGO2017vwq} (orange and grey contours).}
 \label{fig1:M-R}
\end{figure}

It is useful to estimate the effects of EMSG modifications to GR
on the observational properties of neutron stars using three
nuclear EOSs with diverse high-density behaviour.
Figure \ref{fig1:M-R} displays the mass-radius relations obtained as solutions of TOV equations using three different representative 
EoSs: namely the relativistic NL3 \cite{Pais16,Horowitz01,Carriere03} based on the RMF model, the 
relativistic BSR2 \cite{Dhiman07,Agrawal10} which is an extended version of RMF with non-linear meson-meson cross-couplings, and the non-relativistic
 Sly4 \cite{Chabanat98} based on the SHF approach. To explore the EMSG modifications to GR, one
 ensures that the magnitude of the parameter $\alpha$ should be such that it only induces 
 perturbative changes in the structure of NS compared to GR. To this end, we consider the 
 {\it maximum (positive) value} of $\alpha_{\rm max} \approx 10^{-37}$ cm$^3$/erg as estimated
 in Ref. \cite{ABCEK2018} from combined constraints (at the $68\%$ confidence level) from $M-R$
 measurements of NSs in low-mass X-ray binaries \cite{Ozel:2016oaf}.  
 The exact $\alpha_{\rm max}$ value depends on the matter EoS that causes distinct effective stiffening/softening inside the NS. 
 For the NL3, BSR2, Sly4 models the upper-bound on $\alpha$ are respectively (0.9, 1.9, 3.6)$\times 10^{-37}$ cm$^3$/erg.
 For larger values of $\alpha >0$, the mass-radius curves increase continuously and will not satisfy the $M-R$ constraints 
 (see Fig. 3 in Ref. \cite{ABCEK2018}). For clarity of presentation in the figure we have 
 considered a fixed maximum value of $\alpha_{\rm max} \approx 10^{-37}$ cm$^3$/erg which 
 is also consistent for diverse sets of EOSs used in subsequent correlation analysis. 
 The {\it minimum value} of $\alpha <0$ is obtained by the NS conditions: 
 $dm/dr >0$ from the surface ($r=R$) to the center ($r=0$) of the star, $dP/dr<0$ 
 from central pressure $P_c$ to the surface value $P=0$,
 as well as the stability criterion $dM/d\rho_c \geq 0$, where the equality
 criterion provides the maximum mass $M_{\rm max}$ at the central energy density $\rho_c$.
 From the TOV equation \eqref{TOV1}, the $dm/dr >0$ condition is satisfied 
 when $\alpha\rho_c(1+ 8P_c/\rho_c + 3P^2_c/\rho_c^2) > - 1$. On the other hand, Eq. \eqref{TOV2} determines the stability condition $dP/dr<0$ for $\alpha <0$ which is dictated by the dominant last (negative) term within the square brackets leading to the condition
$2\alpha[3P + \rho(\dif\rho/\dif P)^2] > -1$. 
Note that this term itself gives singularity in the TOV equation at a negative 
$\alpha_s = -2[3P + \rho(\dif\rho/\dif P)^2]^{-1}$ 
 which is then avoided  by the $dP/dr<0$ condition. Hence no further restriction on $\alpha < 0$ is required to exclude the singular value in EMSG, in contrast to that in the Palatini $f({\cal R})$ gravity for a conformal equation of state $P = \rho/3$ \cite{Lope-Oter23}.
Figure \ref{fig1:M-R} depicts the mass-radius results with
 the minimum values of $\alpha_{\rm min} = -(1.9, 1.2, 0.8)\times 10^{-37}$ cm$^3$/erg 
 determined using these stability conditions for the (NL3, BSR2, Sly4) EoSs. 
 (However, in the correlation analysis involving several 
 diverse sets of EoSs, we will use the  minimum value of $\alpha_{\rm min} \simeq -10^{-38}$ cm$^3$/erg \cite{ABCEK2018} which can be obtained by inserting, in the stability condition for $dm/dr>0$, the typical (lower) central values $P_c/\rho_c \sim 0.2$ 
and $\rho_c \sim 10^{37}~{\rm erg}^{-1} \rm {cm}^3$.)

\begin{figure}[t]
 \begin{center}
\includegraphics[width=0.48\textwidth,angle=0]{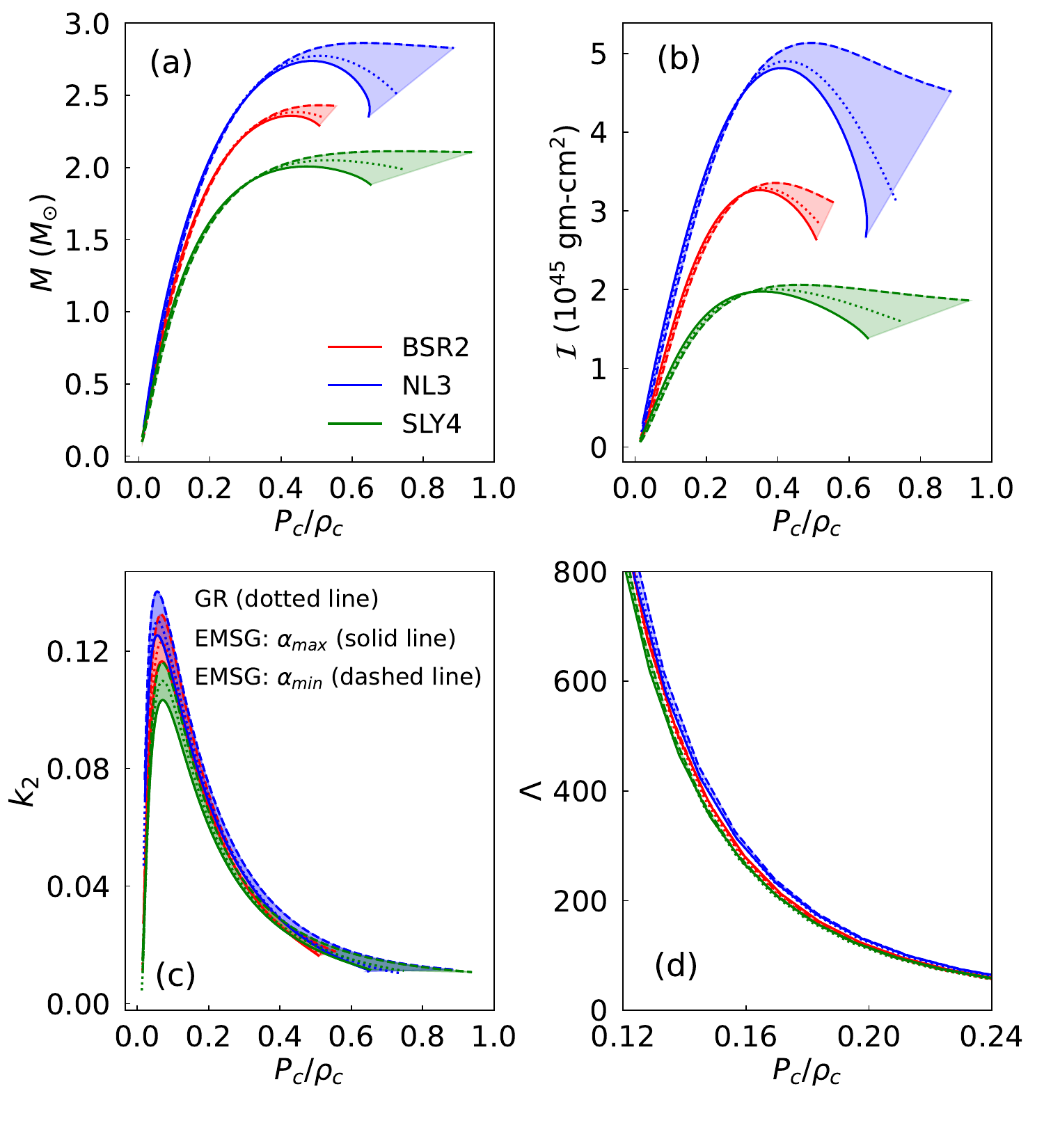}
\end{center}
\vspace{-1.1cm}
  \caption{The ratio $P_c/\rho_c$ for central values of pressure and energy densities as a function 
  of (a) neutron star mass $M$, (b) moment of inertia $I$, (c) Love number $k_2$, 
  (d) tidal deformability $\Lambda$ in the NL3, BSR2, and Sly4 nuclear EoSs. 
  The results are in GR an EMSG with $\alpha$ parameters 
  as given in Fig. \ref{fig1:M-R}.}
 \label{fig2:pe-MIKD}
\end{figure}
\begin{figure*}[t]
 \begin{center}
\includegraphics[width=\textwidth,angle=0]{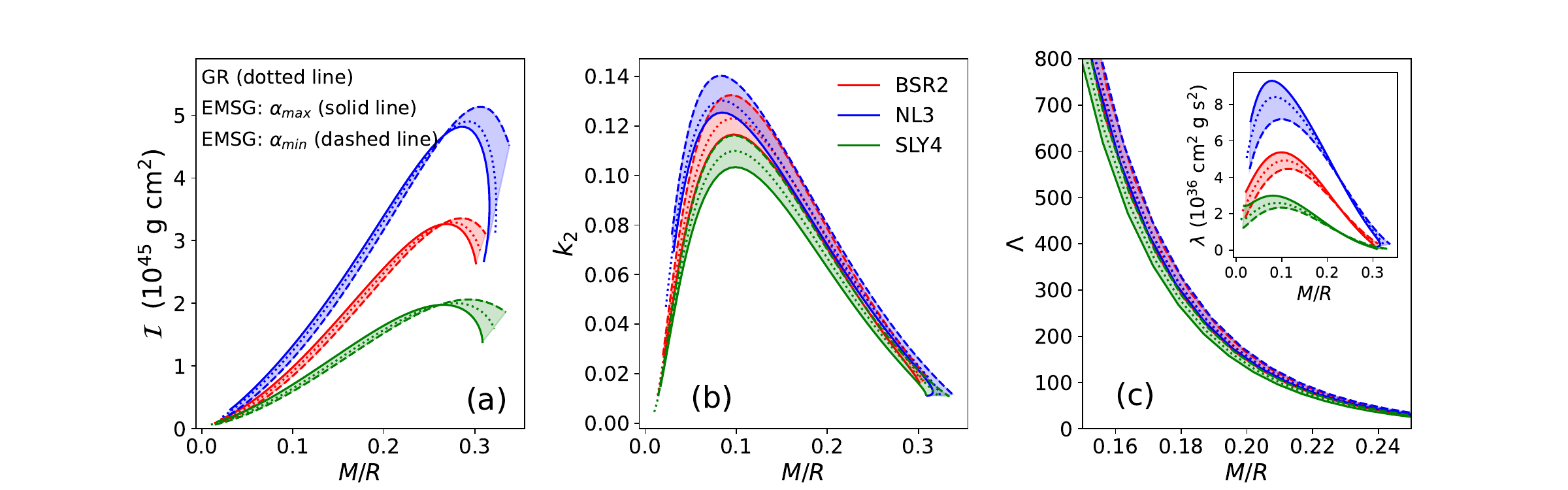}
\end{center}
 \vspace{-0.5cm}
  \caption{Dependence of neutron star compactness parameter $M/R$ on (a) moment of inertia $I$, (b) dimensionless Love number $k_2$, (c) dimensionless tidal deformability $\Lambda$ and quadrupole polarizability $\lambda$ (inset) in the  NL3, BSR2, and Sly4 nuclear EoSs in the GR and EMSG as referred to in Fig \ref{fig1:M-R}.}
 \label{fig3:mass-IKD}
\end{figure*}

For our choice of the three EoSs, the NL3 has the stiffest $P-\rho$ variation and hence reveals the
largest maximum mass $M_{\rm max}$ and the correspondingly the largest radius $R_{\rm max}$. As compared to 
GR, the EMSG model, in general, causes an  effective stiffening of the EoS at low densities and softening at high densities for 
$\alpha > 0$, and conversely 
for $\alpha < 0$. 
The maximum masses are found to remain almost unaffected, whereas the radii increase (decrease) 
appreciably for the maximum (minimum) values of $\alpha$ employed here. 
For the constant positive value of $\alpha\equiv \alpha_{\rm max} \approx 10^{-37}$ cm$^3$/erg, 
the softest Sly4 has as the largest increase in radii $\Delta R$, 
whereas for $\alpha <0$, the stiffest NL3 (with smallest $\alpha_{\rm min} = -1.9\times 10^{-37}$ cm$^3$/erg) 
exhibit the maximum decreases in $\Delta R$.
In fact, the maximum variation of the radius $\Delta R \approx 0.6$ km is seen for the
$M \sim 0.5 M_\odot$ NS, relative to the GR calculation. 
These EMSG modifications can be understood by noting that TOV equations can be 
represented by a single relevant dimensionless quantity $P/\rho$. This translates to the dimensionless 
compactness parameter $C = GM/Rc^2$ of a star given that a larger degenerate pressure $P$ essentially leads to a larger star radius $R$ \cite{Brown00}. As discussed in subsection \ref{sec:hydrostatics}, the 
corresponding ratio in the EMSG model turns out to be 
$P_{\rm EMSG}/\rho_{\rm EMSG} = [1+ 8\rho P/(\rho^2 + 3P^2)]^{-1}$. Finite limits can be placed
at  $P/\rho =0$ (vacuum), $P/\rho =1/3$ (ultra-relativistic Fermi gas: conformal bound) 
and $P/\rho \leq 1$  (causality condition), which translate respectively to 
$P_{\rm EMSG}/\rho_{\rm EMSG} \in (1, 1/3, \leq 1/3)$. This implies from Eqs. \eqref{TOV1}, \eqref{TOV2}
that for $\alpha > 0$, EMSG stiffens the effective EoS below the conformal bound $P/\rho =1/3$ and softens 
the effective EoS above the bound, and conversely for $\alpha <0$.
Figure \ref{fig2:pe-MIKD}(a) illustrates such a variation of 
the ratio $P_c/\rho_c$ in the NS centers with the masses of the
star sequence. Note that for each EoS shown, most of the stars 
in the sequence are confined within $P_c < \rho_c/3$ leading
to stiffening (softening) for positive (negative) values of 
$\alpha$, and correspondingly predict larger (smaller) star radii. On other hand, 
the stars at and near the maximum mass $M_{\rm max}$ are located above 
the conformality bound $P_c/\rho_c =1/3$ and well within 
the causality constraint $P_c /\rho_c \leq 1$.

Profound implications may follow in EMSG theory for negative $\alpha$. 
For example, various parametrizations of the
nuclear EoSs strive to simultaneously describe the observational tidal deformability bound of 
$\Lambda_{1.4} \leq  580$ of a canonical $1.4M_\odot$ NS inferred from GW170817 event \cite{LIGO2017vwq} and the
maximum mass bound $M_{\rm max} \gtrsim 2M_\odot$. The current tension can be effectively addressed
in EMSG (for $\alpha < 0$) that predicts smaller radii (thereby even smaller $\Lambda \sim (R/M)^5$)
for low mass neutron stars but relatively insensitive to the maximum mass of NSs. Furthermore, 
a star of extremely small mass $M = 0.77^{+0.20}_{-0.17}M_\odot$ and radius  
$R = 10.4^{+0.86}_{-0.78}$ km is estimated within the supernova remnant HESS J1731-347
\cite{Doroshenko2022}, which has posed the 
interesting possibility of exotic strange stars. We emphasize that even a pure nucleonic
star, owing to its small radius in the EMSG strong-field gravity, can be an exciting viable alternative.

Figure \ref{fig3:mass-IKD} explores the EMSG effects on the NS observables:
the moment of inertia $I$, tidal Love number $k_2$ and the tidal deformability 
$\Lambda$ as a function of compactness parameter $C=M/R$ for each of the 
NL3, BSR2, Sly4 EoSs. The variation of these observables on the central pressure
to central energy density ratio $P_c/\rho_c$ for the star sequence are shown
in Fig. \ref{fig2:pe-MIKD}(b)-(d).
As discussed above, the dimensionless $C$ naturally translates into a measure of the 
neutron stars pressure and energy at the center via the relation $M/R \sim P_c/\rho_c$.
The moment inertia can be a useful estimate of the EMSG effects since the
dimensional relation $I \propto MR^2$ gives relatively larger ranges from changes in the radius, 
and moreover, the accuracy of radius estimations are largely limited by uncertainties.
Although the moment of inertia depends on the underlying stiffness/softness of the EoS, 
one notices in Fig. \ref{fig3:mass-IKD}(a), small effects of gravity on the variation of moment of 
inertia with the dimensional parameter $M/R$ in any individual nuclear EoS.
This can be traced from Fig. \ref{fig2:pe-MIKD}(b) to the subdued effect of 
EMSG on the ratio $P_c/\rho_c$, irrespective of the choice of nuclear EoS. 

On the other hand, the variation of Love number $k_2$ of \eqref{eq:k2} 
with compactness in Fig. \ref{fig3:mass-IKD}(b)
shows noticeable modifications in EMSG primarily near the $k_2$
peak at $C \approx 0.1$ that corresponds to $M \approx 1M_\odot$. Whereas, $k_2$ is found to be relatively independent of the details 
of the models of gravity as well as the EoSs at small compactness $C \lesssim 0.05$ that is
dominated by the large crustal radii for these small stellar masses.
At large $C \gtrsim 0.25$ near the maximum mass configurations, the values of $k_2$ 
become much smaller than at their $M_{\rm mass}$. Although $k_2$ is seen here 
to be quite sensitive to the EoS, the EMSG modifications to GR at this
strong gravity field regime are however smaller compared to the observed 
spread in $k_2$ for $1M_\odot$ stars.

Tidal fields from inspiraling binary neutron stars induce a quadrupole polarizabality 
$\lambda = (2/3)k_2R^5$ or a dimensional tidal deformability $\Lambda = (2/3)k_2(R/M)^5$,
which may be sensitive to models of gravity due to $R^5$ dependence.
The inset of Fig. \ref{fig3:mass-IKD}(c) depicts a larger variation of
$\lambda$ with $M/R$ in the EMSG, especially near  $1M_\odot$
as seen at the $k_2$-peak. In contrast, a strong
correlation between the dimensionless tidal deformability $\Lambda$ and compactness
(as well as between $\Lambda - P_c/\rho_c$; see Fig. \ref{fig2:pe-MIKD}(d)) appears, irrespective 
of the models of gravity and the three representative EoSs. In fact, such a tight correlation
was found between $\Lambda_{1.4}$ and $R_{1.4}$ for canonical $1.4M_\odot$ 
pure nucleonic stars \cite{Nandi2018} and with nucleon-quark phase transition \cite{Nandi2021} suggesting the 
possibility to constrain the radius and perhaps the symmetry energy $e_{\rm sym}(\rho_0)$.

\subsection{Correlation analysis between neutron star properties and nuclear matter parameters}

In the following we shall explore the possible correlations between the NS observables 
($R$, $I$, $\Lambda$) with the nuclear matter (NM) saturation parameters 
($K_0$, $Q_0$, $M_0$, $J$, $L$, $K_{\rm sym}$) and linear combination of two
NM parameters (such as $K_0+\beta L$, $M_0+\eta L$, $M_0+\zeta K_{\rm sym}$),
and their impact due to the EMSG theory. To facilitate the
correlation study we include all the RMF, SHF and microscopic
models described in section \ref{sec:eos}. Hereafter, we shall employ
the fixed maximum and minimum values of the parameter $\alpha$ estimated in the EMSG theory \cite{ABCEK2018},
$\alpha_{\rm max} = 10^{-37}$ cm$^3$/erg and $\alpha_{\rm min} = -10^{-38}$ cm$^3$/erg 
for all the EoSs employed, which ensure stable configurations for all the neutron stars.

\begin{figure}[t]
 \begin{center}
\includegraphics[width=0.44\textwidth,angle=0]{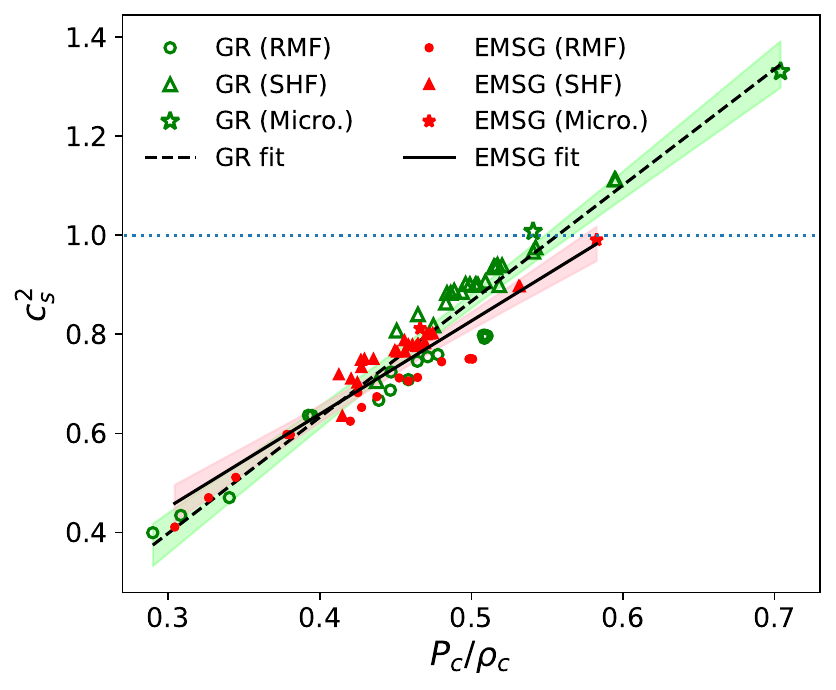}
\vspace{-0.7cm}
\end{center}
  \caption{Correlation between speed of sound squared $c_s^2$ and the ratio of central values of pressure and energy densities $P_c/\rho_c$ corresponding to maximum mass stars in the RMF, SHF and microscopic models of EoS. The results are in EMSG with $\alpha = 10^{-37}$ cm$^3$/erg (red solid symbols) and in GR (green open symbols). The lines represent the linear best-fit and the shaded regions correspond to $95\%$ confidence band.}
\label{fig4:sss-pe}
\end{figure}

Before attempting such NS-NM correlations, it is instructive to employ the causality 
bound of speed of sound squared $c^2_s = dP/d\rho \leq 1$ to impose limits 
on the maximum value $P_c/\rho_c$ at the center and its natural transform $R_{\rm max}/M_{\rm max}$ for the superdense NS matter. 
Figure \ref{fig4:sss-pe} displays the correlation between central sound speed $c_s^2$ and the reduced central pressure 
$\widetilde{P}_c \equiv P_c/\rho_c$  
from all the diverse EoSs in the EMSG theory for the parameter $\alpha = 10^{-37}$ cm$^3$/erg
(red solid symbols) and in GR (green open symbols).
The central speed of sound for the maximum mass stars increases with the reduced central pressure which is a measure of stiffness of dense nuclear matter
inside a NS. Reasonably good correlations are found, given that a broad class of EoSs are employed. Albeit, the correlations are found to be distinct in EMSG and GR which are a direct consequence of the strong-field gravity. Compared to GR, a stiffer EOS in EMSG below the conformality bound $P=\rho/3$ and a 
softer EoS above this bound for $\alpha > 0$  
manifest in an increase in $c_s^2$  at 
$\widetilde{P}_c \lesssim 0.45$ and a reduced 
$c_s^2$ at larger $\widetilde{P}_c$.
We find that the 
conformability bound ($c_s^2 \leq 1/3$) appears to be violated  at the central densities reached in all the stars. Also depicted in Fig. \ref{fig4:sss-pe} are
the linear regressions between $c_s^2$ and $P_c/\rho_c$ 
in EMSG (solid lines) and GR (dashed lines) with slope and intercept as:
\begin{align}\label{eq:cs2Pc-EMSG}
c_s^2 &= (1.880 \pm 0.120)\frac{P_c}{\rho_c}  + (-0.113 \pm 0.054), ~~ [{\rm EMSG}] \\  
c_s^2 &= (2.341 \pm 0.102)\frac{P_c}{\rho_c}  + (-0.304 \pm 0.050), ~~ [{\rm GR}].
\label{eq:cs2Pc-GR}
\end{align} 
The intercept at $c_s^2=1$ enables to set an upper bound on reduced 
central pressure $\widetilde{P}_c$ that is enforced by the causality requirement
$c_s^2 \leq 1$. Our analysis suggests a central upper bound of $\widetilde{P}_c \lesssim 0.592$ 
in EMSG and $\widetilde{P}_c \lesssim 0.557$ in the effectively stiffer EoS in GR.

\begin{figure}[t]
 \begin{center}
 \includegraphics[width=0.45\textwidth,angle=0]{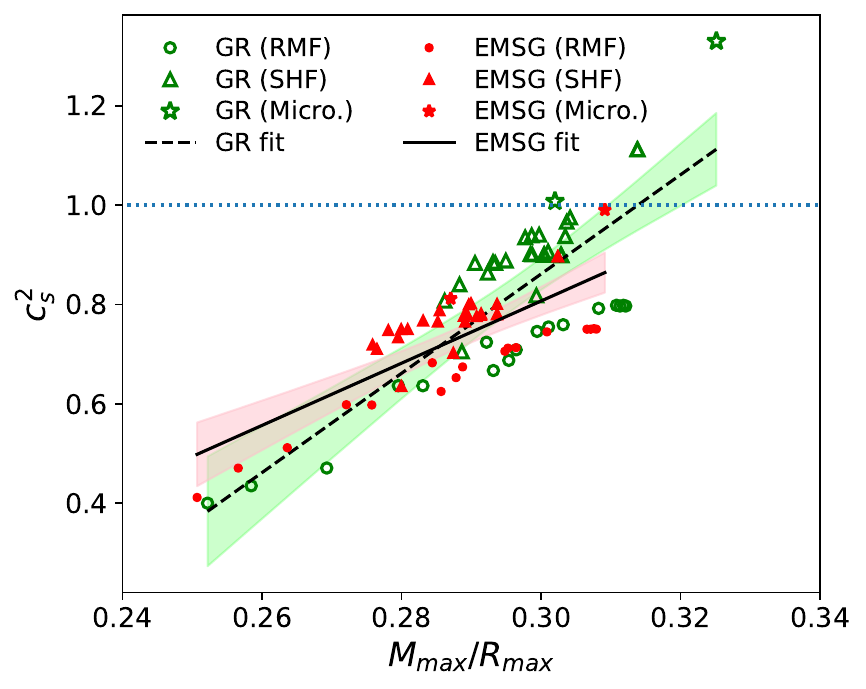}
 \end{center}
 \vspace{-0.7cm}
  \caption{Same as Fig. \ref{fig4:sss-pe} but for correlation between $c_s^2$ and the ratio 
  $M_{\rm max}/R_{\rm max}$ for the maximum mass and corresponding radius of neutron stars.}
  \label{fig5:sss-RM}
\end{figure}
\begin{figure*}[t]
 \begin{center}
 \includegraphics[width=\textwidth,angle=0]{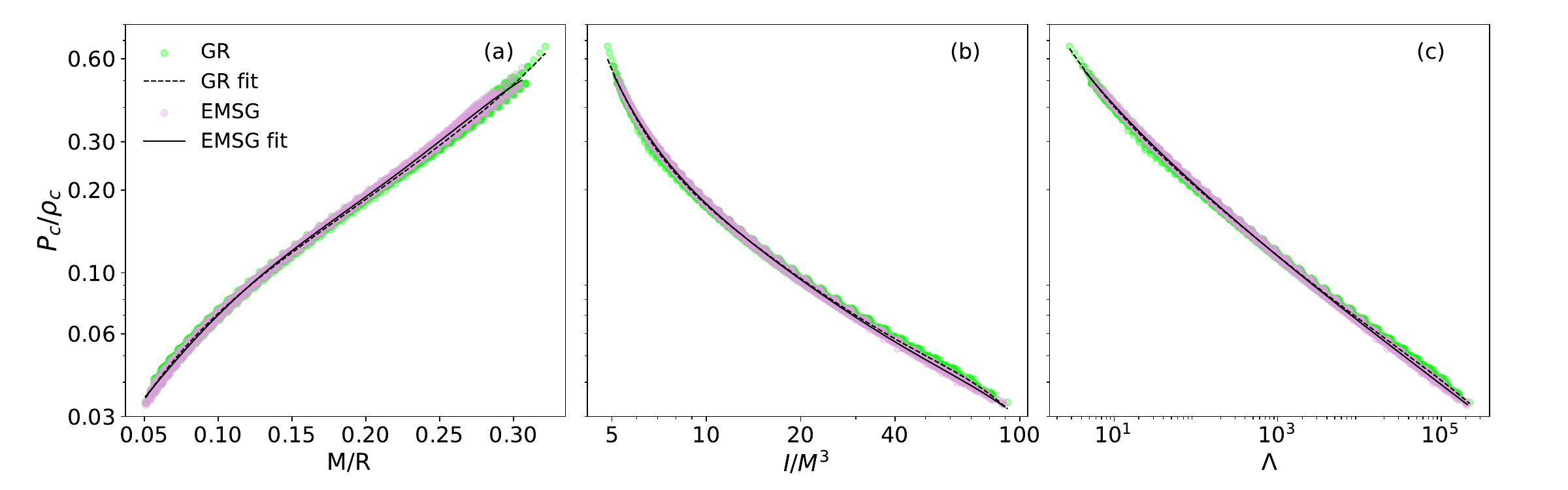}
 \end{center}
 \vspace{-0.7cm}
  \caption{Correlation between $P_c/\rho_c$ and several neutron star quantities: 
  (a) compactness $M/R$, (b) scaled moment of inertia $I/M^3$, 
  (c) tidal deformability in the RMF, SHF, and microscopic models of EoS. The results are 
  in GR (green circles) and EMSG with $\alpha = 10^{-37}$ cm$^3$/erg (magenta stars). 
  The lines are fifth-order polynomial fits to the correlations.}
 \label{fig6:pe-MID}
\end{figure*}

 The dependence of $c_s^2$ on $\widetilde{P}_c$ translates into its dependence on
 $R_{\max}/M_{\rm max}$ for the maximum mass configurations at the NS centers as displayed in \ref{fig5:sss-RM}. The intrinsic structures
 in the TOV equations however prevent a perfect dimensionless mapping leading 
 to some cluttering in the correlation with the compactness parameter. In fact,
 some model-dependence is revealed, viz the relatively stiffer EoSs in the relativistic mean field model generate stars with
 large $M_{\rm max}$ but also have fairly large radius $R_{\rm max}$, and thus yield less 
 compact stars with smaller $c_s^2$ as compared to those in the non-relativistic Skyrme-Hartree-Fock models.
 In general, the central speed of sound is found to increase with the compactness of the NSs \cite{Cai2023pkt}.
 The EMSG theory, that predicts slightly larger $R_{\rm max}$ for $\alpha > 0$, have smaller sound speed compared to GR. Also depicted in Fig. \ref{fig4:sss-pe} are
 our constructed linear regressions between $c_s^2$ and $M_{\rm max}/R_{\rm max}$ with $95\%$ confidence bands by accounting for the EoS scatter in the EMSG and GR as:
\begin{align}\label{eq:cs2MR-EMSG}
c_s^2 &= (6.259 \pm 0.808) \frac{M_{\rm max}}{R_{\rm max}}  + (-1.071 \pm 0.223), ~~ [{\rm EMSG}] \\  
c_s^2 &= (9.995 \pm 1.167) \frac{M_{\rm max}}{R_{\rm max}}  + (-2.3172 \pm 0.347), ~~ [{\rm GR}].
\label{eq:cs2MR-GR}
\end{align} 
 From the causality condition, we obtain a central upper bound on the compactness of about 
 $C_{\rm max} \equiv M_{\rm max}/R_{\rm max} \lesssim 0.338$ that corresponds 
 to a lower limit for the radius $R_{\rm max}/{\rm km} \gtrsim 4.370 M_{\rm max}/M_\odot$
 in EMSG theory. Similarly in the GR, a maximum compactness of $C_{\rm max} \lesssim 0.314$ 
 translates to the radius bound of
 $R_{\rm max}/{\rm km} \gtrsim 4.704 M_{\rm max}/M_\odot$. These compactness
 bounds are much smaller than Buchdahl's upper limit $C_{\rm max}^{\rm up} = 4/9$ \cite{Buchdahl1970}. 
 A direct comparison of our estimated radius bound can be made with the NICER observations 
 for the PSR J0740+6620 \cite{Riley2021} radius of about $12.39^{+1.30}_{-1.98}$ km with 
 a mass $M \approx 2.08^{+0.07}_{-0.07}M_\odot$ and for the PSR J0030+0451 radius 
 $\approx 12.71^{+1.14}_{-1.19}$ km with a mass $\approx 1.34^{+0.15}_{-0.16}M_\odot$ \cite{Riley2019}. 
 Clearly, our estimated radii lower bounds in the models of gravity are well
 consistent with the NICER measurements.
 Inversely, constraints on the EoS variables can be applied by using the NS measurements. 
 For example, the central mass-radius ($M=2.08M_\odot,\: R=12.39 \:{\rm km})$ of PSR J0740+6620 yields 
 from Eq. \eqref{eq:cs2MR-EMSG}, a central sound speed of $c_s^2 \approx 0.481$ which (from Eq. \eqref{eq:cs2Pc-EMSG}) corresponds
 to a reduced central pressure of $\widetilde{P}_c \approx 0.316$.

\begin{figure}[b]
 \begin{center}
\includegraphics[width=0.48\textwidth,angle=0]{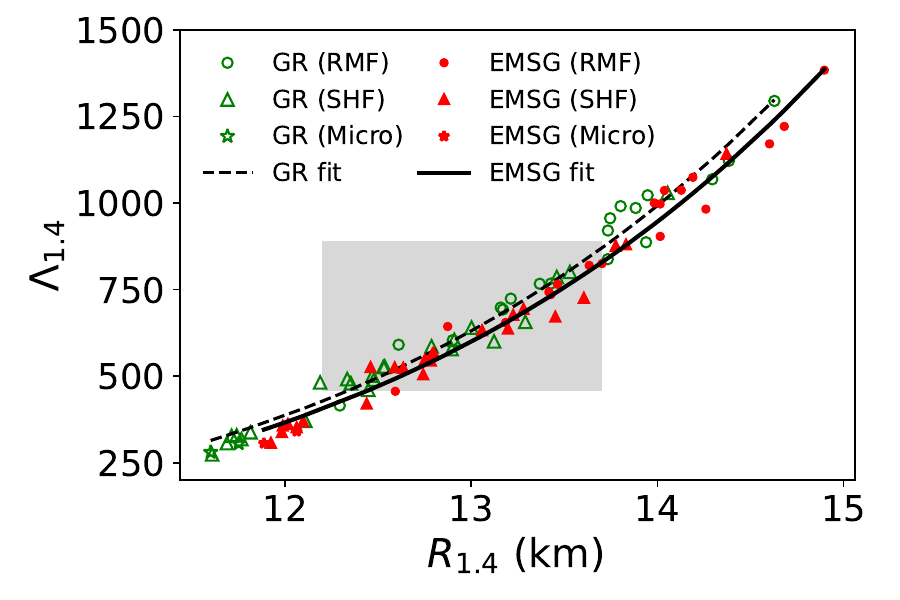}
\vspace{-0.9cm}
\end{center}
\caption{Correlation between tidal deformability $\Lambda_{1.4}$ and radius $R_{1.4}$
of neutron star of mass $M=1.4M_\odot$. The symbols 
are the same as in Fig. \ref{fig4:sss-pe}.
The fits are represented by
$\Lambda_{1.4} = 8.37\times 10^{-5} (R_{1.4}/{\rm km})^{6.15}$ in EMSG (solid line) and
$\Lambda_{1.4} = 9.67\times 10^{-5} (R_{1.4}/{\rm km})^{6.12}$ in GR (dashed line).
The grey shaded region refers to $R_{1.4} = 12.9^{+0.8}_{-0.7}$ km and $\Lambda_{1.4} = 616^{+273}_{-158}$ bounds from GW190814 event \cite{LIGOS2020zkf}.}
\label{fig7:Radius-Lambda}
\end{figure}
\begin{figure*}[t]
 \begin{center}
\includegraphics[width=0.7\textwidth,angle=0]{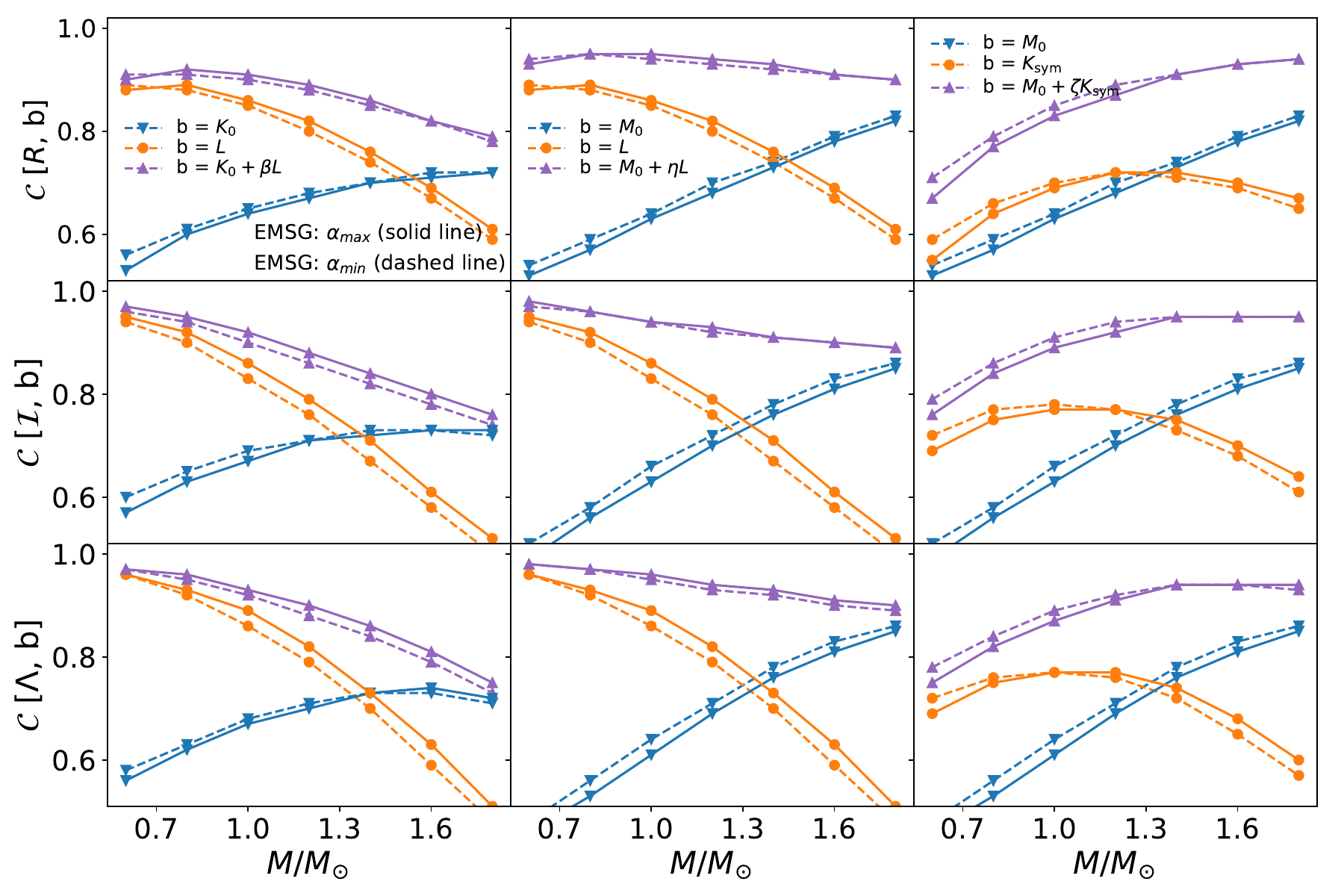}
\end{center}
\vspace{-0.7cm}
\caption{Neutron star mass $M$ dependence of the Pearson correlation coefficients ${\cal C}$ between NS observables and nuclear EoS parameters within the RMF, SHF and microscopic models.
The correlations involve NS radii $R$ (top panels), moment of inertia $I$  
(middle panels), tidal deformability $\Lambda$ (bottom panels), with the
EoS parameters $b \in (K_0, L, M_0, K_{\rm sym})$ and their linear combinations:
$K_0 + \beta L$,  $M_0 + \eta L$, 
$M_0 + \zeta K_{\rm sym}$. The results are in the EMSG gravity model with coupling parameter $\alpha_{\rm min} = -10^{-38}$ cm$^3$/erg (dashed lines) and $\alpha_{\rm max} = 10^{-37}$ cm$^3$/erg (solid lines).}
 \label{fig8:corr}
\end{figure*}

Various approximate universal relations connecting the NS observables, such 
as the compactness $C= M/R$, dimensionless moment of inertia 
$\widetilde{I} \equiv I/M^3 \propto C^{-2}$, dimensionless tidal deformability 
$\Lambda \propto C^{-5}$, have  been established that are insensitive to the 
microscopic details of the high-density EoSs
\cite{Yagi13,Malik18,Carson19,Yang23}. It is useful to test and validate these 
relations with respect to our collection of EoSs and to the models of gravity as well. 
Figure \ref{fig6:pe-MID} shows correlations between the reduced central pressure 
$\widetilde{P}_c \equiv P_c/\rho_c$ with the dimensionless quantities 
$C, ~\widetilde{I}, ~\Lambda$ of the NSs. Remarkably tight correlations do exist 
that are insensitive primarily to the EoSs and the gravitational interactions. 
Measurements of these NS observables thus provide accurate estimates of pressure 
at the fiducial densities which can be invoked in the model EoSs to constrain nuclear 
interactions and the nuclear matter parameters. Polynomial fits up to fifth order 
of the form $\ln \widetilde{P}_c = \sum_{i=0}^5 a_i {\cal S}^i$, where 
${\cal S} \equiv (C, ~\ln\widetilde{I}, ~\ln\Lambda)$ are shown in EMSG and GR. 

To explore the impact of tidal deformability on the structure of a star in EMSG, we display in Fig. \ref{fig7:Radius-Lambda} the correlation between $\Lambda_{1.4}$ and radius $R_{1.4}$ for stars of $M=1.4M_\odot$ computed for all the EOSs.
The increase of $R_{1.4}$ with $\Lambda_{1.4}$ is simply due to the fact that $\Lambda$ quantifies the variation of gravitational field relative to a point-mass object. The proportionality of $\Lambda$ on
$R^5$, reveals in a tight correlation, i.e. an approximate universal relation independent of the input EoSs. Interestingly, the increase in $R$  for positive $\alpha$ values 
(as seen in $M$-$R$ curve of Fig. \ref{fig1:M-R}) enforces distinct class of universalities
for the EMSG and GR gravity models. 
In fact, the correlations can be expressed as $\Lambda_{1.4} = {\cal A} \: R_{1.4}^\xi$, 
that are practically EOS-insensitive and reveal a 
small but finite dependence on the models of gravity, as evident from the parameters 
${\cal A} = 8.37 (9.67) \times 10^{-5}$ and $\xi = 6.15 (6.12)$ for EMSG (GR).
A bound on $\Lambda_{1.4} = 190^{+390}_{-120}$ at $90\%$ confidence was extracted by LIGO-VIRGO 
from the observed binary neutron-star merger GW170817 event using 
Bayesian analysis with a common EoS for the compact binaries \cite{LIGO2018cki}. 

The more recent observation of GW190814 signal \cite{LIGOS2020zkf} from coalescence 
of a massive $(22.2-24.3)M_\odot$ black hole and a compact object of mass 
$(2.50-2.67)M_\odot$ can provide an 
intriguing opportunity to test modifications of GR due to large asymmetry in the masses.
While the primary component of GW190814 is conclusively proven to be a black hole (BH), the lack of measurable tidal
deformations and the nondetection of an electromagnetic counterpart are consistent with the secondary component being either a NS or a BH. Considering that the NSBH scenario cannot be 
completely discounted, a stringent constraint 
was given \cite{LIGOS2020zkf} for the NS that favors a stiff EoS which translates to radius and tidal deformability of a canonical $1.4M_\odot$ NS of
 $R_{1.4} = 12.9^{+0.8}_{-0.7}$ km 
and $\Lambda_{1.4} = 616^{+273}_{-158}$ at $90\%$ credible level. Currently, GW190814
offers this unique observational bound simultaneously for $R_{1.4}$ and $\Lambda_{1.4}$, which we shall use hereafter to constrain the EoS.
Imposing this bounds in Fig. \ref{fig7:Radius-Lambda} (grey shaded region), 
we find that the rather small sensitivity of the gravity models cannot 
be disentangled from the $R_{1.4}-\Lambda_{1.4}$ relation. In this respect, we note that the higher multipole moments of the gravitational signal, that enables to test the multipolar structure of gravity, do not show any deviations from the predictions of GR  \cite{LIGOS2020zkf}. While our derived correlations are consistent with the GW190814 constraints, the bound 
clearly favours EoSs with soft symmetry energy $e_{\rm sym}(\rho)$ at density 
$\rho \sim 2\rho_0$ and rules out the super-stiff EoSs that predict large radii. 

We next analyze the correlation between the neutron star bulk observables presented above with the
key nuclear matter (NM) parameters of the EoS, namely $K_0$, $M_0$, $L$, $K_{\rm sym}$ and 
a few selected linear combinations of these parameters.
In particular, we also explore the influence of the EMSG modifications to GR on these correlations. 
The Pearson correlation coefficient, ${\cal C}[a,b]$, has been used for a quantitative analysis of
a linear correlation between two quantities $a$ and $b$, which can be expressed as \cite{Brandt97},
\begin{equation}
{\cal C}[a,b]=\frac{\sigma_{ab}}{\sqrt{\sigma_{aa}\sigma_{bb}}}\, ,\\
\label{eq:cc}
\end{equation}
where the covariance, $\sigma_{ab}$, is given by
\begin{equation}
\sigma_{ab}=\frac{1}{N_m}\sum_i a_i b_i -\left(\frac{1}{N_m}\sum_i a_i\right
)\left(\frac{1}{N_m}\sum_i b_i\right ) .
\end{equation}
The index $i$ runs over the number of models $N_m$ used in the analysis; 
$a_i$ and $b_i$ respectively refer to the NS properties (such as radius, moment of interta, deformability) 
at a fixed mass, and the NM parameters in the EoS. A correlation coefficient ${\cal C}[a,b] = \pm 1$
would suggest a perfect correlation/anticorrelation between the two quantities of interest,
and ${\cal C}[a,b] = 0$ would indicate no correlation.

\setlength{\tabcolsep}{4.8pt}
\begin{table}[t]
\caption{\label{tab1} Pearson correlation coefficients ${\cal C}$ between 
the parameters in RMF, SHF, microscopic nuclear models and the
the radii $R_{1.4}$, moment of inertia $I_{1.4}$ 
and tidal deformability $\Lambda_{1.4}$ of a $1.4M_\odot$ mass NS.
The nuclear matter EoS parameters are incompressibility $K_0$, skewness $Q_0$, 
slope of incompressibility $M_0$, symmetry energy $J$, its slope $L$, and 
curvature $K_{\rm sym}$, and the parameter $K_{\tau}$ all calculated at the saturation density. 
The correlations are calculated in GR and EMSG with coupling parameter 
$\alpha_{\rm min} = -10^{-38}$ cm$^3$/erg and $\alpha_{\rm max} = 10^{-37}$ cm$^3$/erg,
and denoted respectively by superscript (${\rm GR}, <, >$) on the NS quantities.}
\vspace{-0.5cm}
\begin{center}
\begin{tabular}{cccccccc} 
\hline\hline
   & $K_0$  & $Q_0$ & $M_0$ & $J$ & $L$ & $K_{sym}$ & $K_{\tau}$\\
\hline
$R_{1.4}^< $ & 0.704 &  0.572 &  0.743 &  0.559 &  0.743 &  0.713 & $ -0.686 $ \\
$R_{1.4}^> $ & 0.698 &  0.554 &  0.728 &  0.576 &  0.762 &  0.717 & $ -0.693 $ \\
$R_{1.4}^{\rm GR} $ & 0.703 & 0.570 & 0.742 & 0.560 & 0.744 & 0.713 & $ -0.686 $\\[1mm]
$\Lambda_{1.4}^< $  &  0.730 &   0.605 &   0.778 &  0.507 &   0.696 &   0.719 & $ -0.662 $ \\
$\Lambda_{1.4}^> $  &  0.729 &   0.572 &   0.757 &  0.521 &   0.732 &   0.736 & $ -0.670 $ \\
$\Lambda_{1.4}^{\rm GR} $  &  0.734 & 0.602 & 0.778 & 0.501 & 0.698 & 0.720 & $-0.662 $ \\[1mm]
$I_{1.4}^< $  &  0.729  &  0.609  &  0.781  &  0.451  &  0.673  &  0.731  & $ -0.625  $  \\
$I_{1.4}^> $  &  0.724  &  0.582 &   0.761  &  0.473  &  0.706  &  0.745  & $ -0.635 $ \\
$I_{1.4}^{\rm GR} $  &  0.728 & 0.607 & 0.779 & 0.453 & 0.676 & 0.733 & $-0.626$ \\
 \hline \hline
\end{tabular}
\end{center}
\end{table}

Figure \ref{fig8:corr} displays the NS mass dependence of the Pearson correlation coefficients between the NS quantities ($R, ~I,~\Lambda$) and the EoS parameters
in the EMSG model with coupling parameter $\alpha_{\rm min} = -10^{-38}$ cm$^3$/erg 
(solid lines) and $\alpha_{\rm max} = 10^{-37}$ cm$^3$/erg (dashed lines) 
corresponding to the minimum and maximum estimated bounds \cite{ABCEK2018}. 
Noticeable effects of the parameter $\alpha$ on the correlation coefficients are seen. 
The isovector parameter, $L$, corresponding to the slope of symmetry energy induces somewhat enhanced correlation due to larger radius (smaller compactness $M/R$) for positive $\alpha_{\rm max}$. In contrast, the correlations 
with the isoscalar parameters $K_0, ~M_0$ show opposite dependence on $\alpha$. On the other hand, for the isovector symmetry curvature 
$K_{\rm sym}$, the correlation strengths between the positive and negative $\alpha$ show inversion at $M \approx 1.2M_{\odot}$.
Further, the low mass NSs exhibit much stronger sensitivity to $L, ~K_{\rm sym}$ (characterized by large correlation function) which gradually decreases with increasing NS mass, and eventually at $M \gtrsim 1.4 M_\odot$ the isoscalar parameters $K_0, ~M_0$ dominate the correlations. Such trends can be 
understood form the expressions of pressure on energy density (i.e. the EoS) when these are expressed in term of the NM parameters 
\cite{Alam16}. The linear combinations, $K_0 +\beta L$, $M_0+\eta L$ and $M_0+\zeta K_{\rm sym}$ indicate the strongest sensitivity to
the NS observables over the entire mass range, wherein the correlations are designed to yield optimum values by tuning the coefficients 
$\beta, \eta, \zeta$. This means, that these combinations have a stronger correlation compared to that for the individual nuclear
parameters. Interestingly, the correlation $M_0+\zeta K_{\rm sym}$, which shows an increasing trend with NS masses, 
has the largest value near the canonical $1.4M_\odot$ star in case of all the NS observables $R, ~I,~\Lambda$.
For orientation, we have listed in Tables I and II the correlation coefficients of the NS quantities with the individual
NM parameters and their linear correlations for 1.4 solar mass NS.

\setlength{\tabcolsep}{6pt}
\begin{table}[t]
\caption{\label{tab2} Pearson correlation coefficients ${\cal C}$ between the NS
quantities and linear combinations of EoS parameters in the  
RMF, SHF, and microscopic models. The EMSG parameters and notations are the same as in Table I.} 

\vspace{-0.2cm}
\begin{center}
\begin{tabular}{ccccccc}
\hline\hline
& \multicolumn{2}{c}{$K_0+\beta L$}  & \multicolumn{2}{c}{$M_0+\eta L$}
& \multicolumn{2}{c}{$M_0+\zeta K_{\rm sym}$} \\ 
 \cline{2-3} \cline{4-5} \cline{6-7}
& $\beta$ & ${\cal C}$ & $\eta$ & ${\cal C}$ & $\zeta$ & ${\cal C}$  \\ 
\hline
$R_{1.4}^< $ &  0.878 & 0.848   & 16.972  & 0.923   & 5.351  & 0.914  \\
$R_{1.4}^> $ &  0.963 & 0.858   & 18.484  & 0.926   &  5.596  & 0.907   \\
$R_{1.4}^{\rm GR} $ &  0.886 & 0.849   & 17.103  & 0.923   &  5.373  & 0.914   \\[1.2mm]
$\Lambda_{1.4}^< $ & 0.670 & 0.837    & 13.846  & 0.919   & 5.008  & 0.941  \\
$\Lambda_{1.4}^> $ & 0.768 & 0.856    & 15.979  & 0.925   & 5.482  & 0.937  \\
$\Lambda_{1.4}^{\rm GR} $ & 0.666 & 0.840    & 13.918  & 0.920   & 5.027  & 0.942  \\[1.2mm]
$I_{1.4}^< $  &  0.614 & 0.823   & 12.917  & 0.907   & 5.136  & 0.950 \\
$I_{1.4}^> $  &  0.710 & 0.839   & 14.820  & 0.912   & 5.548  &  0.945 \\
$I_{1.4}^{\rm GR} $  &  0.622 & 0.824   & 13.081  & 0.908   & 5.173  &  0.949 \\
\hline\hline
\end{tabular}
\end{center}
\end{table}

\begin{figure}[t]
 \begin{center}
\includegraphics[width=0.48\textwidth,angle=0]{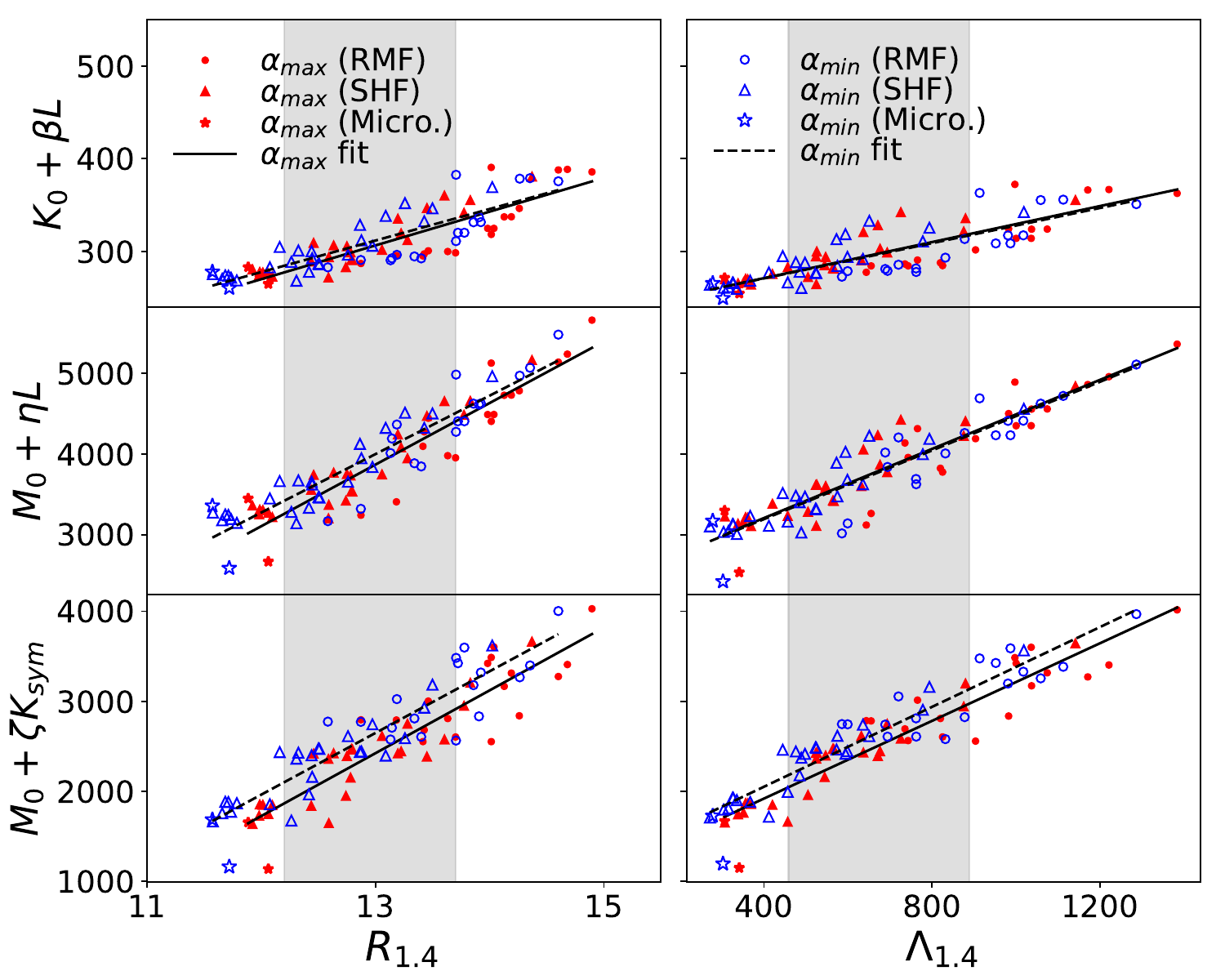}
\end{center}
\vspace{-0.7cm}
\caption{Correlation between NS quantities $R_{\rm 1.4}$, $\Lambda_{1.4}$ and linear combinations of 
nuclear matter parameters $K_0 + \beta L$,  $M_0 + \eta L$, $M_0 + \zeta K_{\rm sym}$ 
in the EMSG gravity model with coupling parameter $\alpha_{\rm min} = -10^{-38}$ cm$^3$/erg (blue open symbols) 
and $\alpha_{\rm max} = 10^{-37}$ cm$^3$/erg (red solid symbols). 
The linear regressions are for $\alpha_{\rm min}$ (dashed lines) and $\alpha_{\rm max}$ (solid lines).
The grey shaded regions refer to $R_{1.4} = 12.9^{+0.8}_{-0.7}$ km and $\Lambda_{1.4} = 616^{+273}_{-158}$ bounds from GW190814 event \cite{LIGOS2020zkf}.}
\label{fig9:corrNS-NM}
\end{figure}

In Fig. \ref{fig9:corrNS-NM} we display the envisaged strong correlation between the NM parameters
$K_0 +\beta L$, $M_0+\eta L$ and $M_0+\zeta K_{\rm sym}$ with the radii $R_{1.4}$ (left panels) and 
tidal deformability $\Lambda_{1.4}$ (right panels) for $1.4M_{\odot}$ NS in the EMSG theory. 
Such strong correlations can be traced essentially to the increase in the NS radii 
with the increase of isocalar and symmetry energy pressures at $\rho \sim (1.5-2)\rho_o$ \cite{Lattimer01}.
Correspondingly, the stiffer effective EoS for $\alpha>0$ at this density range leads to smaller correlation
strength relative to $\alpha <0$ as evident from the linear regression fits to these correlations.
For $\alpha_{\rm max} = 10^{-37}$ cm$^3$/erg, the constructed linear regressions (solid lines) can be represented as:
\begin{align}
&K_0 +\beta L = (36.42\pm 3.36) R_{1.4} + (-167.01 \pm 44.51),  \nonumber \\
&M_0+\eta L  = (762.49 \pm 47.82) R_{1.4} + (-6045.71 \pm 632.56), \nonumber\\
&M_0+\zeta K_{\rm sym}  = (698.48 \pm 50.02) R_{1.4} + (-6656.65 \pm 661.64), \nonumber \\
&K_0 +\beta L  = (0.10 \pm 0.01) \Lambda_{1.4} + (232.44 \pm 6.80),  \nonumber\\
&M_0+\eta L   = (2.14 \pm 0.13) \Lambda_{1.4} + (2352.40 \pm 101.82),  \nonumber\\
&M_0+\zeta K_{\rm sym}  = (2.16 \pm 0.12) \Lambda_{1.4} + (1057.95 \pm 93.13). 
\end{align}
Here ($K_0, L, M_0, K_{\rm sym}$) are in the units of MeV and $R_{1.4}$ in km.
We use the above set of relations (and the coefficients $\beta, \eta, \zeta$
listed in Table II)
in conjunction with the GW190814 bound on $R_{1.4} = 12.9^{+0.8}_{-0.7}$ km and $\Lambda_{1.4} = 616^{+273}_{-158}$ \cite{LIGOS2020zkf}
to estimate the nuclear matter parameters. We utilize the quite accurately
constrained nuclear matter incompressibility at the saturation density of 
$K_0 = 240 \pm 20$  obtained from analysis of isoscalar giant monopole resonance (ISGMR)
collective excitations in $^{90}$Zr and $^{208}$Pb nuclei \cite{Colo2004,Todd-Rutel05,Colo2013}. 
The central values are estimated to be
($L =65.22,\: M_0 = 2585.15, \: K_{\rm sym} = - 41.35$) MeV for the $R_{1.4}$ constraint,
and ($L=70.36, \: M_0 = 2546.36, \: K_{\rm sym} = - 28.79$) MeV for the $\Lambda_{1.4}$ constraint.
The obtained slope of symmetry energy are in line with $L=(50.0 \pm 15.5)$ MeV extracted from
available nuclear masses of heavy nuclei \cite{Fan14}, as well as the reported values of
$L=(106 \pm 37)$ MeV \cite{Reed2021} and $L=(54 \pm 8)$ MeV \cite{Reinhard2021} from analysis of neutron skin thickness measurements of $^{208}$Pb by the PREX-II experiment. Further, our estimated slope of the incompressibility $M_0$ are consistent with the empirical constraint $M_0 = (1800-2400)$ MeV determined
by comparing Skyrme-like energy density functional and the energies of the ISGMR  
$^{132}$Sn and $^{208}$Pb nuclei \cite{Khan12,Khan13}. Our estimate of the curvature parameter of symmetry energy lies well within the present fiducial value 
$K_{\rm sym} = - 107 \pm 88$ MeV obtained from combined analysis of NS observables of GW170817 signal
\cite{Malik18,Tsang04,Carson19}, energy density functionals constrained by terrestrial experiments and observational data \cite{Mondal16}, and metamodeling of nuclear EoS with these constraints \cite{Margueron02}.

Likewise, the linear regressions for correlations with EMSG parameter value $\alpha_{\rm min} = -10^{-38}$ cm$^3$/erg (dashed lines in Fig. \ref{fig9:corrNS-NM}) can be utilized to extract the NM parameters from the GW190814 constraints. The deviations from $\alpha_{\rm max}$ are found to be at the level of about $18\%$ for $L$ and $\sim 32\%$ for $M_0$. The symmetry energy curvature $K_{\rm sym}$ was found to have large sensitivity to $\alpha$ parameter.
Combining all these results, our estimated central values
are found to be $77.88 \lesssim L \lesssim 65.22$ MeV, 
$1951.32 \lesssim M_0 \lesssim 2589.12$ MeV, and
$-41.35 \lesssim K_{\rm sym} \lesssim 117.49$ MeV.
Although the EMSG theory suggests different classes of approximate
relations for various $\alpha$, the nuclear matter parameters obtained
are within the current bounds from various model analysis of terrestrial and observational measurements.

For orientation, the linear regression in general relativity is given as  
\begin{align}
&K_0 +\beta L = (33.93\pm 3.39) R_{1.4} + (-129.60 \pm 44.04),  \nonumber \\
&M_0+\eta L  = (707.10 \pm 46.11) R_{1.4} + (-5196.25 \pm 598.15), \nonumber\\
&M_0+\zeta K_{\rm sym}  = (661.74 \pm 45.28) R_{1.4} + (-5961.00 \pm 587.45), \nonumber \\
&K_0 +\beta L  = (0.09 \pm 0.01) \Lambda_{1.4} + (233.55 \pm 6.80),  \nonumber\\
&M_0+\eta L   = (2.05 \pm 0.13) \Lambda_{1.4} + (2393.63 \pm 94.16),  \nonumber\\
&M_0+\zeta K_{\rm sym}  = (2.14 \pm 0.11) \Lambda_{1.4} + (1218.85 \pm 77.44). 
\end{align}
Adopting the same approach as in EMSG, we obtain in GR the central values 
($L =76.86,\: M_0 = 2610.80, \: K_{\rm sym} = - 6.58$) MeV for the $R_{1.4}$ constraint,
and ($L=73.56, \: M_0 = 2632.62, \: K_{\rm sym} = - 19.00$) MeV for the $\Lambda_{1.4}$ constraint. The estimated NM parameters in GR and EMSG (with $\alpha_{\rm max}$ for example) are found to be somewhat different.
However, the large uncertainty associated with the current bounds (from nuclear experiments and/or detected NS observables) prevents a clear preference for the various descriptions of gravity.

We would like to mention that the above bounds in EMSG theory of 
gravity are obtained by employing the observational constraints
that are often performed based on general relativity. In particular, 
the tidal deformability is extracted from the gravitational wave signal using 
waveform models derived assuming GR. 
However, the current detectors are not sensitive to the new class of modes that may appear from additional degrees of freedom in the 
modified theories of gravity which do not exist in GR \cite{Mendes:2018qwo}. Consequently, the
detected observables ($M,~R,~\Lambda$) cannot clearly resolve GR from the modified theories of gravity.

While our results are based on EoSs with purely nucleonic degrees of freedom, 
possible existence of exotic phases, such as phase transition to hyperons \cite{Schaffner-Bielich:2000nft}, 
kaon condensation \cite{Pons:2000iy}, deconfined quark matter \cite{Nandi2021,Sen22} at high densities in the core of neutron stars 
cannot be excluded. In general, the onset of phase transition inside the star at nucleon density $\rho > \rho_0$ 
softens the EoS resulting in decrease in the speed of sound, masses, radii, moment of inertia 
and tidal deformability of NSs. Hence, as compared to GR results, the reduction in the magnitude of these observables
as found here in EMSG theory for $\alpha > 0$, will be further accentuated with phase transition. Conversely, the
increase seen for $\alpha < 0$ will be minimised by softening of the EoS in presence of phase transition.
The validity of the universal relations connecting $P_c/\rho_c$ with compactness, moment of inertia and tidal
deformability as seen in EMSG and GR (Fig. \ref{fig6:pe-MID}) is expected to persist even with phase transition, as 
universality has been established with phase transition in GR itself \cite{Yagi:2016bkt}. Since non-nucleonic degrees of freedom 
(for realistic EoS) are likely to appear in massive stars only, one expects the strong correlations observed between 
the nuclear matter parameters and low mass neutron star observables (Figs. \ref{fig8:corr}, \ref{fig9:corrNS-NM}) 
will have marginal effect.

\section{Conclusions}
\label{sec:conclusion}

Using a representative set of accurately calibrated models of nuclear equations of state, 
we have investigated within the energy-momentum squared gravity theory 
(EMSG: a non-minimal matter-coupling extension to general relativity) 
the impact of strong-field gravity on several properties of dense neutron star. In particular,
correlations between nuclear matter parameters at saturation density and the neutron
star observables were studied in the EMSG theory to ascertain the effectiveness of the
theory and to quantify its modifications to the predictions in general relativity (GR). By using three
realistic EoSs (NL3, BSR2, Sly4), we first showed that for a {\it fixed value} of 
the coupling strength $\alpha$ in EMSG, the NS mass-radius curves are affected differently 
as compared to GR predictions. The softest EoS in Sly4 enforces the largest increase in the radii 
of especially the small mass stars for the positive $\alpha$ value. Whereas,
the stability conditions, $dm/dr >0$ and $dP/dr <0$ (from center to surface of NS), 
enable smallest $\alpha <0$ in the stiffer NL3 EoS and correspondingly provides the largest
decrease in the radii near the $1M_\odot$ neutron star. While the variation of the NS compactness $C=M/R$ with the
moment of inertia and tidal deformability in particular, are quite small,
the peak value of tidal Love number $k_2$ was found to have appreciable modifications 
to GR in the EMSG model.

We next explored the correlations between the NS observables and nuclear matter EoS
parameters in EMSG and GR.
An approximate universal correlation, independent of the nuclear model EOSs,  was established
for the variation of the central speed of sound squared $c_s^2$  with the reduced pressure
$\widetilde{P}_c \equiv P_c/\rho_c$ and its natural transform the compactness  
$C_{\rm max} \equiv M_{\rm max}/R_{\rm max}$ at the center of the stars. We found that 
$c_s^2$ has a linear increase with $\widetilde{P}_c$ and $C_{\rm max}$. However, the universality 
is violated to some extent by the strong-field gravity that induces distinct correlations for different 
values of the parameter $\alpha$ in EMSG. For instance, the causality bound
on the NS mass-radius curve in  EMSG suggested a lower limit on the star radius
$R_{\rm max}/{\rm km} \gtrsim 4.37 M_{\rm max}/M_\odot$ in direct contrast to 
the GR bound of $R_{\rm max}/{\rm km} \gtrsim 4.70 M_{\rm max}/M_\odot$. 

We also demonstrated that gravity modifications have marginal
effects on the universal properties of the compactness, ${\widetilde I}$, tidal deformability relations with the reduced central pressure  and thus $\widetilde{P}_c$ could be inferred from measurements of NS properties.
A truly universal correlation within the realms of current observational bounds was found 
between the measurable radius and tidal deformability of $1.4M_\odot$ NS that is practically
Eos-insensitive and depicted marginal separation between different classes of universality 
in the EMSG and GR. 
The correlation between nuclear matter incompressibility $K_0$,
its slope $M_0$, the symmetry energy slope $L$ and curvature $K_{\rm sym}$ and their linear combinations revealed the previously
studied correlation with NS radii and deformability. These relations were found to be quite sensitive to EMSG theory of gravity that may hinder a precise estimation of the nuclear matter parameters from these correlations. To conclude,
our study emphasizes that certain neutron star observables are insensitive to nuclear EoSs and gravity modifications and can be employed as approximate universal relations to determine the EoS parameters, whereas small, yet detectable signatures of gravity effects are evident in some neutron star observables.

\section*{ACKNOWLEDGMENTS}
N.A. and S.P. acknowledge financial support by the Department of Atomic Energy (Government
of India) under Project Identification No. RTI 4002.

\end{document}